\documentclass[sigconf]{acmart}
\usepackage{cleveref}

\usepackage{bm}
\usepackage{bbm}
\usepackage{booktabs,subcaption,amsfonts,dcolumn}
\usepackage{wasysym}
\usepackage{hhline}
\usepackage{float}
\usepackage{commath}

 \usepackage{multirow}
 \usepackage[ruled,vlined,linesnumbered]{algorithm2e}
\usepackage{booktabs,siunitx}
\newcommand{\rstep}[1]{\textcolor{blue}{Step #1}}
\usepackage{stmaryrd}
\usepackage{trimclip}
\usepackage{pifont}
\usepackage{enumitem}
\usepackage{balance}

\AtBeginDocument{%
  \providecommand\BibTeX{{%
    \normalfont B\kern-0.5em{\scshape i\kern-0.25em b}\kern-0.8em\TeX}}}

\copyrightyear{2023}
\acmYear{2023}
\setcopyright{rightsretained}
\acmConference[SIGIR-AP '23]{Annual International ACM SIGIR Conference on
Research and Development in Information Retrieval in the Asia Pacific
Region}{November 26--28, 2023}{Beijing, China}
\acmBooktitle{Annual International ACM SIGIR Conference on Research and
Development in Information Retrieval in the Asia Pacific Region (SIGIR-AP '23),
November 26--28, 2023, Beijing, China}\acmDOI{10.1145/3624918.3625313}
\acmISBN{979-8-4007-0408-6/23/11}

\begin{document}

\title{Vertical Allocation-based Fair Exposure Amortizing in Ranking}

\author{Tao Yang}

\affiliation{%
   \institution{University of Utah}
     \streetaddress{50 Central Campus Dr.}
   \city{Salt Lake City}
   \state{Utah}
   \country{USA}
   \postcode{84112}
}
\email{taoyang@cs.utah.edu}
\author{Zhichao Xu}

\affiliation{%
   \institution{University of Utah}
     \streetaddress{50 Central Campus Dr.}
   \city{Salt Lake City}
   \state{Utah}
   \country{USA}
   \postcode{84112}
}
\email{zhichao.xu@utah.edu}

\author{Qingyao Ai}
\affiliation{%
   \institution{DCST,  Tsinghua University \\ Quan Cheng Laboratory, Zhongguancun Laboratory}
   \city{Beijing}
   \country{China}
   \postcode{100084}
}
\authornote{Corresponding authors}
\email{aiqy@tsinghua.edu.cn}

\begin{abstract}
Result ranking often affects consumer satisfaction as well as the amount of exposure each item receives in the ranking services.
Myopically maximizing customer satisfaction by ranking items only according to relevance will lead to unfair distribution of exposure for items, followed by unfair opportunities and economic gains for item producers/providers. 
Such unfairness will force providers to leave the system and discourage new providers from coming in. 
Eventually, fewer purchase options would be left for consumers, and the utilities of both consumers and providers would be harmed. Thus, to maintain a balance between ranking relevance and fairness is crucial for both parties. 
In this paper, we focus on the exposure fairness in ranking services.
We demonstrate that existing methods for amortized fairness optimization could be suboptimal in terms of fairness-relevance tradeoff because they fail to utilize the prior knowledge of consumers.
We further propose a novel algorithm named \textbf{Ver}tical Allocation-based \textbf{Fair} Exposure Amortizing in Ranking, or \textbf{VerFair}, to reach a better balance between exposure fairness and ranking performance.
Extensive experiments on three real-world datasets show that VerFair significantly outperforms state-of-the-art fair ranking algorithms in fairness-performance trade-offs from both the individual level and the group level.

\end{abstract}






\maketitle
\section{Introduction}
Ranking techniques have been extensively studied and applied across online marketplaces (e.g. e-commerce websites, such as Amazon) and social media (e.g. suggested people to follow on Twitter/TikTok). 
Traditionally, the focus of ranking models is to maximize \textit{consumer-side satisfication}, or \textit{relevance}.
The ranklists presented to the consumers are usually constructed by sorting the candidate items according to the estimated relevance of the consumer-item pair.
However, some recent studies \cite{biega2018equity,singh2018fairness} have revealed that this consumer-centered strategy will allocate most of the exposure to few top-ranked popular items and their providers (e.g. products/sellers on e-commerce websites and content/content creators on media platforms), which often referred to as the Winners-Take-All phenomenon. 
Since exposure directly influences opinion (e.g. ideological orientation of presented news articles) or economic gain (e.g. revenue from item sales or streaming), the unbalanced distribution of exposure will eventually drive the other less popular items out of the platform while discouraging new items to come in, where few options left for consumers. 
At the end of the day, the utility of the platform, consumers, and providers will all be harmed. 
Therefore, how to allocate exposure to items fairly, or in other words, to guarantee \textit{Provider-side Utility}\footnote{we use provider-side utility, provider-side fairness, and item fairness interchangeably} is crucial in ranking.


Recently, several provider-side fairness definitions \cite{biega2018equity,singh2018fairness,abdollahpouri2020multistakeholder,abdollahpouri2017controlling,zhu2018fairness,liu2018personalizing,karako2018using} have been proposed by the community. 
One of the most well-recognized principles is \textit{Amortized Fairness} \cite{biega2018equity,singh2018fairness}, which hypothesizes that fairness can be reached if the items' exposure distribution could match their relevance distribution. 
Specifically, amortized fairness in ranking is defined from both the individual level and the group level. 
Individual-level amortized fairness considers each result candidate separately. 
If an item is twice as relevant as another item, it should get twice the exposure as well.
Similarly, group-level amortized fairness considers result candidates in groups (e.g., items can be grouped by their brands). 
If the accumulated relevance of one group is twice that of another group, the group should get twice the exposure as well. 


The key research focus of amortized fairness is mainly on \textit{how to reach a good balance between ranking relevance and fairness}. 
While purely ranking according to relevance could harm ranking systems in the long term, purely considering fairness may also sacrifice consumers' utility \cite{patro2020fairrec}, hurt consumers' experience, and eventually drive consumers away from the platform. 
To balance ranking relevance and fairness, existing methods \cite{morik2020controlling,biega2018equity} proposed to dynamically mitigate unfairness in an online manner where they assume no prior knowledge of the customers is available. 
However, this assumption can be suboptimal in terms of fairness-relevance tradeoff~\cite{wu2021twosided} in scenarios where the system does possess some customer information and can pre-compute result rankings for consumers before the serving time, such as email advertisement and e-commerce recommendation.


In this work, we focus on the problem of fairness-relevance balance in ranking and propose a novel algorithm named \textbf{Ver}tical Allocation-based \textbf{Fair} Exposure Amortizing in Ranking (\textbf{VerFair}).
Compared to existing amortized exposure algorithms, VerFair could achieve a better balance between ranking performance and fairness constraints.
VerFair is a post-processing method that does not depend on  any specific relevance estimation model and, therefore can be seamlessly integrated into existing ranking applications.
While most of the existing fairness ranking methods use a horizontal allocation (details in \S\ref{sec:example}) paradigm to allocate items to customers, we propose a novel vertical allocation paradigm that can put more relevant items at the top ranks while still maintaining the fairness of result rankings. 
Based on the vertical allocation paradigm, we subsequently introduce a mechanism to guarantee minimum relevance-induced exposure for each item, given a predefined tolerance of unfairness. 
Through extensive experiments, we demonstrate the proposed method significantly outperforms the state-of-the-art fair method in terms of top ranks' relevance while the minimum relevance-induced exposures of items are still guaranteed.
To summarize, our contributions are three-fold:

\vspace{3pt}
\noindent
\ding{202} We propose a novel post-processing amortized fairness method VerFair that can provably achieve fairness in both group level and individual level for ranking.

\vspace{3pt}
\noindent
\ding{203} We additionally introduce a novel mechanism to guarantee a minimum relevance-induced exposure for all items/item groups.

\vspace{3pt}
\noindent
\ding{204} Extensive experiments demonstrate that VerFair can reach a significantly better balance between fairness and  top ranks' relevance compared to existing exposure-amortized algorithms.
\vspace{3pt}

\section{Related work}
\label{sec:related}

\textbf{\textit{Fairness.}} 
With development of ML techniques, researchers have been interested in the fairness issue brought by them and the corresponding social impacts \cite{hardt2016equality,feldman2015certifying,kamishima2012fairness,zhang2019faht,kearns2018preventing,udeshi2018automated}.
\textit{Fairness in Ranking} \cite{yang2023marginal,ekstrand2023overview,gao2022fair,saito2022fair,wu2022joint,bigdeli2022gender,naghiaei2022cpfair,usunier2022fast,xu2023p} has attracted much attention as ranking plays an important role in modern Internet services, including E-commerce websites and social media platforms.
Given that ranking is a two-sided market, with customers on one side and item providers on another side, we need to consider customers' satisfaction as well as a fair environment for providers \cite{lambrecht2019algorithmic,edelman2017racial,serbos2017fairness,abdollahpouri2017recommender,abdollahpouri2020multistakeholder,gomez2021winner,singh2021fairness,jia2021calibrating,do2021two,biswas2021toward}.
While existing definitions of fairness in ranking vary a lot (refer to \cite{abdollahpouri2017recommender,pitoura2021fairness,raj2022measuring,ge2022explainable} for a comprehensive survey), in this work we focus on amortized exposure fairness.
Existing works on amortized fairness \cite{yang2021maximizing,zehlike2017fa,celis2017ranking,patro2020fairrec} mainly focus on allocating exposure for each item whose relevance has been estimated.
\citet{patro2020fairrec} propose an allocation method named FairRec to achieve fairness, which guarantees an equal frequency for all items in ranklists. 
However, such frequency-based fairness \cite{patro2022fair} ignores the fact that there exists a large skew in the distribution of exposure for different ranks such as position bias \cite{guo2009efficient,dupret2008user,joachims2017accurately}. 


\textbf{\textit{Amortized Fairness.}} 
In this paper, we focus on how to achieve amortized fairness~\cite{biega2018equity,singh2018fairness,pitoura2021fairness} in a post-processing manner. 
In Table \ref{tab:method_comparison}, we make a comparison between several amortized fairness methods which we include as baselines.
Specifically, let's consider a ranking task where there are $m$ users, $n$ items, and length of ranklist for each user is $k$.
\citet{biega2018equity} proposed to carry out $m$ rounds integer linear program (ILP) with $n^2$ decision variables in each round to amortize exposure. 
Since the size of decision variables is a bottleneck for ILP solvers, \citet{biega2018equity} proposed a down-sampling step that helps to reduce the size of candidate sets in each round and there are $O(k^2)$ decision variables in each round. 
Instead of trying to amortize exposure dynamically, \citet{singh2018fairness} adopt Linear Programming (LP) with $n^2$ decision variables to give a static probabilistic ranking, which is mostly infeasible, given a large number of items. 
Besides, LP methods assume one single relevance distribution for items, while the ILP method can work with multiple relevance distributions. 
Thus, the ILP method is more suitable in ranking system given that the distribution of personal relevance varies from person to person. 
Besides linear programming, \citet{morik2020controlling} propose a more efficient fair ranking algorithm, FairCo,  which first determines each item's unfairness and then boosts ranking score of under-exposed items with a proportional controller.
Unlike the post\-processing method mentioned above, methods such as PG-Rank \cite{singh2019policy}, MMF \cite{yang2021maximizing}, PLRank~\cite{oosterhuis2021computationally}, MCFair~\cite{yang2023marginal}, and FARA~\cite{yang2023fara} opt to achieve amortized fairness in learning to rank procedure. 
\citet{wu2021tfrom} provide a theoretically analysis of the relationship between ranking relevance and fairness. 
One additional note is that, in this paper, the fairness we are considering is different from learning representations to achieve a fair model where relevance rating should be independent of some
sensitive attribute \cite{zhu2018fairness,beutel2019fairness,abdollahpouri2017controlling}.

\setlength{\textfloatsep}{0pt}%
\begin{table}[t]
    \centering
    \caption{Comparison of different amortized fairness methods. Attributes include whether they can achieve amortized fairness, work with personal relevance or not, need to do down sampling or not, and their computational complexity. FairRec is not an amortized fairness method, but we still include it here for completeness.}
    \vspace{-5pt}
    \resizebox{0.85\textwidth}{!}{\begin{minipage}{\textwidth}

    \begin{tabular}{c p{1.5cm} p{1.5cm} p{1.5cm} p{1.5cm}}\toprule
    \multirow{3}{*}{Method} &\multicolumn{4}{c}{Attributes}\\\cline{2-5}
    & Amortized fairness & Personal relevance & Down sampling & Compu. complex. \\ \hline
    LP \cite{singh2018fairness}& \multicolumn{1}{c}{\ding{51}} & \multicolumn{1}{c}{\ding{55}} & \multicolumn{1}{c}{\ding{55}} & High \\
    ILP \cite{biega2018equity}& \multicolumn{1}{c}{\ding{51}} & \multicolumn{1}{c}{\ding{51}} & \multicolumn{1}{c}{\ding{51}} & Medium \\ 
    FairCo \cite{morik2020controlling}& \multicolumn{1}{c}{\ding{51}} & \multicolumn{1}{c}{\ding{51}} & \multicolumn{1}{c}{\ding{55}} & Low \\
    FairRec \cite{patro2020fairrec}& \multicolumn{1}{c}{\ding{55}} & \multicolumn{1}{c}{\ding{51}} & \multicolumn{1}{c}{\ding{55}} & Low \\
    VerFair (Ours)& \multicolumn{1}{c}{\ding{51}} & \multicolumn{1}{c}{\ding{51}} & \multicolumn{1}{c}{\ding{51}} & Low \\ \bottomrule 
    \end{tabular}
    \end{minipage}}
    \label{tab:method_comparison}
\end{table}
\vspace{0pt}
\section{Background and Prior Knowledge}
\label{sec:background}
In this section, we will introduce related definitions. A summary of notations used in this paper is shown in Table \ref{tab:notation}.
\begin{table}[t]
	\caption{A summary of notations.}
        \vspace{-5pt}
	\small
	\def\arraystretch{1}
	\resizebox{\columnwidth}{!}{
	\begin{tabular}
		{| p{0.12\textwidth} | p{0.32\textwidth}|} \hline
		$m, n, k, k_c$ & The number of consumers $m$, the number of items $n$, ranklist length $k$, cutoff of NDCG evaluation $k_c$.\\\hline
		$d, \mathcal{D}, u$, $\mathcal{U}, G,\mathcal{G},G_{index}$ & The item set $\mathcal{D}$, an item $d$, the consumer set $\mathcal{U}$, a consumer $u$, a group $G$, the group set $\mathcal{G}$. $G_{index}(d)$ is a function which returns $d$'s group. \\\hline
		$R(d,u), R(d)$, $E(d)$, $E(G), R(G)$ & item $d$'s personal relevance $R(d,u)$ to consumer $u$, item $d$'s averaged relevance $R(d)$ across all consumers, item $d$'s accumulated exposure  $E(d)$, group $G$'s accumulated exposure  $E(G)$, group $G$'s accumulated relevance $R(G)$\\\hline
	\end{tabular}\label{tab:notation}
	}
\vspace{5pt}
\end{table}

\noindent
\textbf{\textbullet\ \textit{Exposure and Fairness}}.
To optimize ranking fairness, there are two concepts that are of importance: relevance and exposure. 
Personal relevance $R(d,u)$ indicates preference of consumer $u$ toward an item $d$. 
Aside from personal relevance, average relevance $R(d)$ is also widely used in ranking fairness \cite{morik2020controlling}. Expected average relevance $R(d)$ or global relevance indicates global preference for all consumers toward item $d$. It is defined by marginalizing personal relevance:
\begin{equation}
\vspace{-2pt}
    R(d)=\int_{u\in \mathcal{U}}prob.(u)R(d,u) 
\vspace{0pt}
\end{equation}
where $\mathcal{U}$ is the set of all consumers, $prob.(u)$ is consumer $u$'s probability in $\mathcal{U}$. In this paper, we refer to average relevance as \textit{relevance} unless otherwise explicitly specified. 

In existing works \cite{morik2020controlling,singh2018fairness,biega2018equity}, exposure is defined as the examination probability, or in other words, how likely an item will be viewed in a ranked list. 
Previous studies mostly model the item exposure following the Position Bias Assumption~\cite{joachims2017accurately,craswell2008experimental} where examination probability depends on the rank position in a ranklist.
As an item could be at different ranks in different consumers' ranklists, we compute the accumulated exposure an item $d$ gets by:
\begin{equation}
    E(d)=\sum_{\pi \in \mathcal{B}} p_{\textit{rnk}(d|\pi)}
    \label{eq:eq(d)}
\end{equation}
where $\mathcal{B}$ denotes all ranklists, $\textit{rnk}(d|\pi)$ is the rank of item $d$ in ranklist $\pi$ and $p_{\textit{rnk}(d|\pi)}$ is the examination probability of item $d$ in the ranklist $\pi$. 

Aside from above definition of relevance and exposure for individual item $d$, we could also define group-level relevance and exposure over items of the same brand, or from the same producer.
In the group level,
we can accumulate relevance $R$ and exposure $E$ for items within group $G$ respectively,
\begin{equation}
\vspace{-2pt}
\begin{split}
    R(G)=\sum_{d\in G}R(d),\;\;
    E(G)=\sum_{d\in G}E(d) 
    \label{eq:RE(G)}
\end{split}
\vspace{-2pt}
\end{equation}

\noindent
\textbf{\textbullet\ \textit{Two-side Utility Measurement}}.
Ranking is a two-sided market, with consumers on one side and providers on the other side. Consumers care more about ranking relevance while providers care more about fairness \cite{wu2021twosided,abdollahpouri2020multistakeholder,abdollahpouri2017recommender}.

\noindent
(1) \textbf{Ranking Relevance}:
Here we use NDCG \cite{jarvelin2002cumulated}, a widely adopted ranking metric to measure ranking relevance from consumer side. Specifically, we use $\textit{NDCG}@k_c$, where $k_c$ is the cutoff position of the ranklist. Note that $\textit{NDCG}@k_c$ is bounded within $[0,1]$.

\noindent
(2) \textbf{Amortized Fairness}:
Fairness is used to measure the ranking quality from the provider side. For amortized fairness, most studies evaluate fairness by measuring the distance between empirical distributions of exposure $E$ and relevance $R$. For example, \citet{biega2018equity} use the $L1$ distance. Here we choose to adopt Jensen–Shannon divergence instead of $L1$ distance since it is bounded to the same range as NDCG, i.e., $[0,1]$, which is better for model comparison and result visualization.
In both individual level and group level, we define the fairness as, 
\begin{equation}
\begin{split}
    \vspace{-2pt}
    \textit{Individual}\: \textit{Fairness}&=1-\textit{JSD}(E(d)||R(d)|d\in \mathcal{D}) \\
    \textit{Group}\: \textit{Fairness}&=1-\textit{JSD}((E(G)||(R(G)|G\in \mathcal{G})
    \end{split}
    \label{eq:fairness}
    \vspace{-2pt}
\end{equation}
where \textit{JSD} denotes Jensen–Shannon divergence which gives the divergence between exposure distribution $E$ and relevance distribution $R$ among all items/item groups. 
Fairness is within $[0,1]$. 
Higher divergence means more unfair ranklists for the providers.


\section{Our Method}
\label{sec:verfair_algo}
In this section, we introduce VerFair, an algorithm for amortized fairness. 
We start this section by extending the discussion of amortized fairness with \textit{Exposure Quota} (\S\ref{sec:quota}), 
followed by two motivating examples to help readers understand the concept of vertical allocation and how VerFair could guarantee the items' exposure quota (\S \ref{sec:example}) while reaching better ranking relevance at top ranks. 
We illustrate the details of VerFair (\S \ref{sec:algo}) and provide theoretical proof to VerFair's exposure quota guarantee (\S \ref{sec:proof}).
\subsection{Exposure Quota}
\label{sec:quota}
Here, we consider a ranking task, where we need to select and rank $k$ unique items to each consumer, 
and there are in total $m$ consumers and $n$ candidate items.
For this ranking task, the total exposure $E_{\textit{total}}$ is fixed:
\begin{equation}
\vspace{-2pt}
    E_{\textit{total}}=\sum_{i=1}^{m}\sum_{j=1}^{k}p_{i,j}
    \label{eq:sum all rank}
\vspace{-2pt}
\end{equation}
where $p_{i,j}$ is the examination probability of consumer $i$ towards rank $j$. And the exposure of all items/item groups  should sum to the total exposure,
\begin{equation}
\vspace{-5pt}
    \sum_{d\in \mathcal{D}}E(d) \equiv \sum_{G\in \mathcal{G}}E(G) \equiv E_{\textit{total}}
    \label{eq:sum exp const}
\vspace{-1pt}
\end{equation}
Then the key question is how to distribute this total exposure to each item/group fairly. 
To tackle this, we define $\textit{Quota}(d|\alpha)$ as the fair share of exposure for an item $d$ as:
\begin{equation}
\vspace{-2pt}
    \textit{Quota}(d|\alpha)=\frac{ R(d) \times \alpha \times  E_{\textit{total}} }{\sum_{d'\in \mathcal{D}}R(d') }
    \label{eq:quota_ind}
\vspace{-2pt}
\end{equation}
where $\mathcal{D}$ is the set of candidate items and $\alpha$ indicates the fraction of total exposure to be allocated for fairness.
Here, we require the exposure that item $d$ actually gets (i.e., $E(d)$) should be greater than or equal to its fair share of exposure:
\begin{equation}
\vspace{-2pt}
    E(d) \geq \textit{Quota}(d|\alpha) \: \:\forall d \in \mathcal{D}
    \label{eq:quota_geq_ind}
\vspace{-2pt}
\end{equation}
where $Quota(d|\alpha)$, the fair share of exposure, can be viewed as relevance-induced minimum exposure. 
Similarly, at group level, the fair share of exposure for group $G$ is:
\begin{equation}
\vspace{-2pt}
     \textit{Quota}(G|\alpha)=\frac{ R(G) \times \alpha \times  E_{\textit{total}} }{\sum_{G'\in \mathcal{G}}R(G') }  
     \label{eq:quota_group}
\vspace{-2pt}
\end{equation} 
A minimum fair exposure for group $G$ is guaranteed if:
\begin{equation}
\vspace{-2pt}
   E(G) \geq \textit{Quota}(G|\alpha) \: \:\forall G \in \mathcal{G}
   \label{eq:quota_geq_group}
\vspace{-2pt}
\end{equation}
When $\alpha =1.0$, i.e., all exposure are used for fair exposure allocation, considering \Cref{eq:sum all rank,eq:sum exp const,eq:quota_ind}, we have equality constraints:
\begin{equation}
\vspace{-2pt}
    \sum_dE(d)\equiv\sum_d \textit{Quota}(d|\alpha=1)
\vspace{-2pt}
\label{eq:strictFairness}
\end{equation}
Inequality constraints defined in Eq. (\ref{eq:quota_geq_ind}) and Eq. (\ref{eq:quota_geq_group}) will degenerate to the above equality constraints when $\alpha =1.0$. 
The equality constraints are the exact constraints of amortized fairness defined in \cite{biega2018equity,singh2018fairness}, where an item's exposure should be proportional to its relevance. 
When $0\le\alpha \le1.0$, $\alpha$  not only decides $\textit{Quota}(d|\alpha)$ (i.e. the minimum fair exposure), but also decides the degree of fairness we try to guarantee. 
An additional note is that our goal is not to find a $\alpha$ that maximize ranking relevance or fairness; 
instead, our goal is to maximize ranking relevance given the same degree of fairness. 
We achieve this by requiring the exposure of each item to satisfy constraints in Eq. (\ref{eq:quota_geq_ind}) and Eq. (\ref{eq:quota_geq_group}). 
We will illustrate more in the rest of this section. 
\subsection{Motivating Examples.}
\label{sec:example}
\subsubsection{Vertical allocation and horizontal allocation.}
\label{sec:verticalBetter}
\begin{table}[t]
    \centering 
    \caption{Ranking performance comparison between vertical allocation and horizontal allocation.}
    \vspace{-5pt}
    \label{tab:example two ways}
    \subcaption{Example consumer-item pair relevance matrix}
    \vspace{-5pt}
    \resizebox{0.75\columnwidth}{!}{
    \begin{tabular}{c c c c}\toprule
    Consumers  &\multicolumn{3}{c}{Item Relevance}\\ \cline{2-4}
        & Item A & Item B & Item C \\ \hline
        Consumer1 & 0.90 & 0.70 & 0.60\\ 
        Consumer2 & 0.55 & 0.70 & 0.90\\ 
        Consumer3 & 0.65 & 0.70 & 0.60\\ 
    \hline
    Average\_relevance& 0.70&0.70&0.70\\ \bottomrule
    \label{tab:rating matrix two ways}
    \end{tabular}
    }
    \vspace{-5pt}
    \subcaption{Ranking performance comparison between vertical and horizontal allocation, where we rank $k=2$ items for all three consumers according to relevance matrix in Table \ref{tab:rating matrix two ways}. We assume constant exposure (=1) for all ranks and the total exposure is 6. We guarantee all ($\alpha=1.0$) of the total exposure to be fair. All items are of the same relevance and each one of them should get exposure of 2 for fairness. 
    (Step) indicate time steps. Vertical allocation can help to get a higher NDCG at rank@1 (top ranks).}
    \vspace{-5pt}
    \resizebox{0.99\columnwidth}{!}{
    \begin{tabular}{l l l l l l l} \toprule
    &\multicolumn{5}{c}{Ranklists}\\ \cline{2-6}
    &\multicolumn{2}{c}{Vertical allocation}  & &\multicolumn{2}{c}{Horizontal allocation}\\  \cline{2-3} \cline{5-6}
    & $1^{st} Rank$&$2^{nd} Rank$&& $1^{st} Rank$&$2^{nd} Rank$\\ 

    \hhline{~--~--} 
        Consumer1& (\rstep{1})A&(\rstep{4})B& &(\rstep{1})A&(\rstep{2})B\\ 
        Consumer2&(\rstep{2})C&(\rstep{5})A& &(\rstep{3})C&(\rstep{4})B\\ 
        Consumer3&(\rstep{3})B&(\rstep{6})C&& (\rstep{5})A&(\rstep{6})C\\ \hhline{~--~--} 
        & NDCG@1& NDCG@2&&  NDCG@1& NDCG@2\\ 
        NDCG&1.000& 0.984 &&0.989&0.989   \\ \toprule
    \end{tabular}
    }
    \label{tab:recommendation}
\end{table}

We give a motivating example in Table \ref{tab:example two ways} before introducing our ranking method. 
In this example, there are 3 consumers and 3 items. 
Table \ref{tab:rating matrix two ways} shows the consumer-item pair relevance. 
Our ranking task is to construct ranklists of length 2 for each consumer given the relevance matrix. 
We follow the amortized fairness principle to construct the ranklists, where items of similar relevance should get similar exposure. 
Since items A, B, and C have the same averaged relevance, they should get the same exposure. 
For simplicity, we assume that the exposure of each rank position is the same and equals 1 for each consumer. 
Thus the total exposure is 6 and each item should fairly show twice. 
Shown in Table \ref{tab:recommendation}, there are two ways to allocate items to consumers, vertical allocation and horizontal allocation. 
In vertical allocation, we fill up one rank for all consumers and then move to the next rank; in horizontal allocation, we fill up all ranks for one consumer and move to the next consumer. 
In this example, we can't choose item B at step 5 in vertical allocation because it has been used up in previous steps. 
The similar situation applies to step 5 in horizontal allocation.

In general, vertical allocation achieves higher NDCG at top ranks, as shown in Table \ref{tab:recommendation}.
Compared to horizontal, vertical allocation have fewer conflicts in allocating relevant items to top ranks and thus higher NDCG at top ranks (see Sec.~\ref{sec:VerBetterNDCG} for theoretical analysis). 
In contrast, we expect horizontal allocation to have a higher long list NDCG since there are more relevant items available for lower ranks.
Previous methods \cite{singh2018fairness,morik2020controlling,biega2018equity,singh2019policy} mostly adopt the horizontal allocation, while we choose vertical allocation due to its superior performance at top ranks. 
Vertical allocation assumes that consumers' information is already known. 
Such assumption is reasonable for certain ranking tasks including email advertisement (ranklists are constructed to all consumers once at the same time) and offline recommendation \cite{patro2022fair,patro2020fairrec}.

\begin{table}[t]
    \centering 
    \caption{Minimal exposure guarantee comparison.}
    \label{tab:toy start pts rec list}
    \vspace{-5pt}
    \subcaption{Example customer-item pair relevance matrix}
    \vspace{-5pt}
    \resizebox{0.75\columnwidth}{!}{
        \begin{tabular}{c c c c} \toprule
        Consumers &\multicolumn{3}{c}{Item Relevance}\\ \cline{2-4}
            & Item A & Item B & Item C \\ \hline
            Consumer1& 0.90&0.80&0.70\\ 
            Consumer2& 0.90& 0.60& 0.80\\ 
            Consumer3& 0.60&1.00&0.90\\ 
            \hline
            Average\_relevance& 0.80&0.80&0.80\\ \toprule
            \label{tab:rating matrix two start}
        \end{tabular}
    }
    \vspace{-5pt}
    \subcaption{
    Comparison of minimal exposure guarantee between starting from origin and from anchor, where we rank $k=2$ items according to relevance matrix defined in Table \ref{tab:rating matrix two start}. 
    \textcolor{red}{*} indicates the anchor point. 
    We assume constant exposure (=1) and guarantee half ($\alpha$= 0.5) of the total exposure to be fair, where each item should get at least exposure of 1.
    We carry out the allocation phase (steps 1-3), then appending phase (steps 4-6), and finally re-sorting phase; $\shortleftarrow$ and  $\shortrightarrow$ indicate moving an item to a higher rank (forward) and to lower rank (backward) in the re-sorting phase. 
    Direction indicates the resorting direction of fairly allocated items (steps 1-3). 
    }
    \vspace{-5pt}
    \resizebox{0.99\columnwidth}{!}{
    \begin{tabular}{l l r l l r} \toprule
    &\multicolumn{5}{c}{Ranklists}\\ \cline{2-6}
    &\multicolumn{2}{c}{Start from origin} & & \multicolumn{2}{c}{Start from anchor} \\  \cline{2-3} \cline{4-6}
    & $1^{st} Rank$&$2^{nd} Rank$&& $1^{st} Rank$&$2^{nd} Rank$\\ 
    \hhline{~--~--} 
    Consumer1& (\rstep{1})A&(\rstep{4})B& &(\rstep{4})B$\shortrightarrow$&$\shortleftarrow$(\textcolor{red}{*}\rstep{1})A\\ 
    Consumer2&(\rstep{2})C$\shortrightarrow$&$\shortleftarrow$(\rstep{5})A&&(\rstep{5})A&(\rstep{2})C\\ 
    Consumer3&(\rstep{3})B&(\rstep{6})C&& (\rstep {6})C$\shortrightarrow$&$\shortleftarrow$(\rstep{3})B\\\hline
    Direction: & \textit{Backward} &&& & \textit{Forward}   \\ \bottomrule
    \label{tab:toy example start list}
    \end{tabular}
    }
    \vspace{-10pt}
\end{table}

\subsubsection{How to guarantee a minimal exposure?}
\label{sec:minimal exp}
We show a sample usage of vanilla ranking strategy in left side of Table \ref{tab:toy example start list}. 
In the allocation phase (step 1-3), the vertical allocation starts from $(1,1)$ and moves to $(2,1)$, $(3,1)$, after which each item's minimum exposure (quota) is met. 
In the appending phase (step 4-6), the allocation algorithms fill the rest parts of the ranklist. 
In the re-sorting phase, for consumer 2, item C is moved to the $2^{nd}$ rank, because item A has higher relevance for consumer 2. 
In practice, as exposure drops from higher ranks to lower ranks, this may lead to the reduced exposure of item C, breaking the minimum exposure guarantee. 
We argue that the break of minimum exposure guarantee in vanilla ranking strategy is due to the fact that the items that satisfies the minimum exposure requirements in the allocation phase can only be moved to the lower part of the ranklists in the re-sorting phase, leading to their reduced exposure. 
Then what if we can make the items that satisfies the minimum exposure requirements be only moved to the higher part of the ranklists in the re-sorting phase? 
In right side of Table \ref{tab:toy example start list}, the allocation phase (step 1-3) starts from $(1,2)$ and moves to $(2,2)$, $(3,2)$. 
In the appending phase (4-6), similarly, the rest parts of the ranklists are filled. 
In the re-sorting phase, for consumer 1, item A is moved to rank $1^{st}$ because of its higher relevance. And similar re-sort is performed for consumer 3 and item B. 
In this case, the items that satisfies the minimum exposure requirements can only be moved to the higher part of the ranklists, and the minimum exposure guarantee stays intact.
The latter ranking strategy is different from the vanilla one because its allocation phase starts from the middle of the ranklist, while the vanilla strategy starts from the top.

Formally, we introduce the definition of \textbf{Anchor Point}: \textit{instead of starting from the first customer and the first rank, the vertical allocation starts from $\hat{c}^{th}$ consumer and $\hat{r}^{th}$ rank, or \textit{anchor point} $(\hat{c},\hat{r})$}. As we can observe from the example above, the usage of anchor point guarantees the minimum exposure requirements in the vertical allocation algorithms. We cover the detailed algorithm to locate the anchor point in \S\ref{sec:det_anchor}.

\subsection{VerFair: Algorithm for Amortized Fairness}
\label{sec:algo}
\setlength{\textfloatsep}{0pt}

In this section, we formally propose a fair ranking algorithm which starts from the anchor point to perform vertical allocation.
The algorithm can reach both individual fairness and group fairness, which are denoted as \textbf{VerFair(Ind)} and \textbf{VerFair(Group)} respectively. 
Since individual-level method (i.e., VerFair(Ind)) can be viewed as a special case of group-level method (i.e., VerFair(Group)) when treating each individual item as a unique group, we provide the VerFair(Group) algorithm in Algo. \ref{algo:fair_exp_group}.
We first introduce how to determine the anchor point, then we walk through three phases of VerFair(Group), namely, the Allocation Phase, the Appending Phase and the Re-sorting Phase.

\subsubsection{Determination of Anchor Point.} 
\label{sec:det_anchor}
As discussed in the motivation example in \S\ref{sec:minimal exp}, the anchor point will help guarantee a minimal exposure. 
In this section, we provide the detailed algorithm to find the anchor point in Algo. \ref{algo:find_anchor}. 
To search the anchor point, we start from the last consumer's last rank, i.e. $(m,k)$, and move vertically backwards towards the first consumer's first rank, i.e. $(1,1)$. The search path sequentially includes, $(m,k), (m-1,k), \dots, (1,k), (m,k-1), (m-1,k-1)\dots,(1,k-1),(m,k-2)\dots$. 
The search procedure stops when the accumulated exposure quota is met. Formally, the search stops at $(\hat{c},\hat{r})$ when
\begin{equation}
\vspace{-2pt}
    E_{(\hat{c},\hat{r})} < \alpha \times E_{total} \leq E_{(\hat{c}-1,\hat{r})}
\vspace{-2pt}
\end{equation}
\begin{equation}
\vspace{-2pt}
    E_{(\hat{c},\hat{r})} = \sum_{j=\hat{r}+1}^{k} \sum_{i=1}^m p_{i,j} + \sum_{i=\hat{c}}^m p_{i, \hat{r}}
    \label{eq:ecr}
\vspace{-2pt}
\end{equation}
where $\alpha$ denotes the fraction of fair exposure in total exposure, $m$ is the number of consumers, $k$ is length of each ranklist, and $p_{i,k-j}$ is the exposure (examination probability) of consumer $i$ towards rank $k-j$. 
The first part of the right side of Eq. (\ref{eq:ecr}) denotes the total exposure from rank $r+1^{th}$ to rank $k$ across all consumers; while the second part of the right side is the total exposure from consumer $\hat{c}+1$ to consumer $m$ at rank $\hat{r}$. 
As the search proceeds, there is a point $(\hat{c},\hat{r})$ where total exposure before $(\hat{c},\hat{r})$ is less than the exposure quota, and the exposure $E_{(\hat{c}-1,\hat{r})}$ for next point $(\hat{c}-1,\hat{r})$ is greater than or equal to the exposure quota.

\begin{algorithm}[t]
\textbf{Input}: 
Set of consumers $\mathcal{U}=[m]$, set of distinct items $\mathcal{D}=[n]$, ranklist size $k$, and the relevance scores $R(d,u)$. Fair exposure fraction
$\alpha\in$\ [0,1]\;
Randomly shuffle $\mathcal{U}$\;
Initialize set of feasible items $F(u)\shortleftarrow \mathcal{D}$ for each $u \in \mathcal{U}$\; 
Initialize ranklists $B(u)\shortleftarrow$ $[\emptyset]_{\times k}$  for each $u \in \mathcal{U}$\;
Initialize $Quota(G)$ according to Eq. (\ref{eq:quota_group}) for $G \in \mathcal{G}$\;
Initialize $\Tilde{E}(G)=0\: \forall G \in \mathcal{G}$\;

Get the anchor point $(\hat{c},\hat{r})$ according to Algorithm (\ref{algo:find_anchor})\;
$(c,r)\shortleftarrow (\hat{c},\hat{r}$)  \;
\textbf{Allocation phase}\;
\While{r$\leq$k}{
    \While{c$\leq$m}{
        $H=\{d|Quota(G_{index}(d))-\Tilde{E}(G_{index}(d))\geq p_{c,r}\}$\;
        $Candidate\_Set \shortleftarrow \{d| H\cap
        F(u_c)$\}\;
        \uIf{$Candidate\_Set \neq  \emptyset
        $}{$Candidate\_Set\shortleftarrow F(u_c)$;\}}
       $ d^*\shortleftarrow\underset{d\in Candidate\_Set}{\arg\!\max} {R(d|u_c)}$\;
       Update $B(u_c)[r]\shortleftarrow d^*$ and $c \shortleftarrow c+1$\; 
       $p_{c,r}\shortleftarrow$ exposure of consumer $c$ at rank $r$ \;
       Update $\Tilde{E}(G_{index}(d^*))=\Tilde{E}(G_{index}(d^*))+p_{c,r}$\;
       Update $F(u_c)\shortleftarrow F(u_c) \backslash d^*$}
       Update $c \shortleftarrow1$ and $r \shortleftarrow r+1$\;
}
\For {$c \shortleftarrow 1$ to $m$}{\For {$r \shortleftarrow 1$ to $k$}{
\uIf{$B(u_c)[r]==\emptyset$}{
$ d^*\shortleftarrow\underset{d\in F(u_c)}{\arg\!\max} {R(d|u_c)}$\;
Update $B(u_c)[r] \shortleftarrow d^*$\; 
Update $F(u_c)\shortleftarrow F(u_c) \backslash d^*$. \textbf{Appending phase}\;} 
}$B(u_c) \shortleftarrow sort(B(u_c))$. \textbf{Re-sorting phase}\; 
}
\textbf{Output}: $B(u)$  for each $u \in \mathcal{U}$\;  
\caption{VerFair(Group)}
\label{algo:fair_exp_group}
\vspace{-3pt}
\end{algorithm}
\vspace{0pt}
\begin{algorithm}[t]
\vspace{-3pt}
\textbf{Input}: exposure quota $\alpha \times E_{total}$, ranklists of  size $m\times k$\; 
Initialize $exposure=0$\;
{\For {$j \shortleftarrow k$ to $1$}{
\For {$i \shortleftarrow m$ to $1$}{
    Update $\textit{exposure} \shortleftarrow \textit{exposure} + p_{i, j}$\;
    \uIf{$\textit{exposure} \geq \alpha \times E_{\textit{total}}$}{
        $break$\;
    }
}
\uIf{$\textit{exposure} \geq \alpha \times E_{\textit{total}}$}{
    $\textit{break}$\;
}
}
}
\textbf{Output}: anchor point $(i, j)$\;  
\caption{Algorithm to find anchor point}
\label{algo:find_anchor}
\vspace{-2pt}
\end{algorithm}

\subsubsection{The Allocation Phase.}
As discussed in the motivation example in \S\ref{sec:verticalBetter}, vertical allocation can help get better ranking relevance at top ranks (theoretical analysis in Sec.~\ref{sec:VerBetterNDCG}). 
Here we discuss how to apply vertical allocation in our algorithm. 
In this algorithm,  we first randomly shuffle the given user set $\mathcal{U}$. Then we initialize the feasible set $F(u)$ for all consumers, which means that all items are available for each consumer at the beginning.
Another empty ranklist $B(u)$ of length $k$ is also created.  Then we use $\Tilde{E}(G)$ (initialize as 0) to store group $G$'s actual allocated exposure.
During the allocation phase, the set $H$ stores items whose group still have exposure quota left with margin $p_{c,r}$, i.e. $Quota(G_{index}(d))-\Tilde{E}(G_{index}(d))\geq p_{c,r}$ with some margin $p_{c,r}$.  $G_{\textit{index}}(d)$ returns the group name. 

$F(u)$ stores items that haven't been selected for consumer $u$. The final available set, i.e. \textit{Candidate\_Set} is an intersection between $H$ and $F(u)$.
Starting from anchor point $(\hat{c},\hat{r})$, i.e. $c^{th}$ consumer and $r^{th}$ rank, we select the most relevant items in $Candidate\_Set$ to fill the ranklists $B$. 
If \textit{Candidate\_Set} is already empty because quota has been used up, \textit{Quota} is no longer under consideration. 
We guarantee the minimum exposure constraint in Eq. (\ref{eq:quota_geq_ind}) in this phase and provide theoretical proof in \S\ref{sec:proof_minimum}.

\subsubsection{The Appending Phase.}
As discussed in the motivation example of \S\ref{sec:minimal exp}, there exists much empty space after the allocation phase. 
In the appending phase, for each consumer $u$, we fill the the empty spaces on $B(u)$ with the most relevant items from the items that are not in her current ranklist, i.e., from the feasible set. 
Note that the selection is no longer constrained by the fair exposure requirement, thus it is purely based on relevance.

\subsubsection{The Re-sorting Phase.}
After the appending phase, for each consumer, her ranklist $B(u)$ is full. 
We need to re-sort each consumer's ranklist according to personal relevance since it is not sorted according to relevance (example in \S\ref{sec:minimal exp}). 
After re-sorting, items selected in the allocation phase will only be moved to higher ranks of her ranklist, as shown in \S\ref{sec:minimal exp}, and we assume the exposure will not drop when the item is moved from lower ranks to higher ranks. 
Thus the minimal exposure guarantee still stays intact after the re-sorting phase. 
Although VerFair is an offline method, it can also be extended to online ranking setting. For example, we can limit the set of consumers to only the active consumers at a certain timestamp. Here we leave this to future work.



\subsection{Theoretical Analysis}
\label{sec:proof}

\subsubsection{Analysis for better NDCG at top ranks}.
\label{sec:VerBetterNDCG}
For simplicity of analysis, we adopt non-personal relevance $R(d)$ and use DCG instead of NDCG (normalized DCG) to analysis this problem. The average $DCG@k_c$ is defined as
\begin{equation*}
\vspace{-2pt}
\begin{split}
    &\textit{DCG}@k_c=\frac{1}{m}\sum_{i=1}^m \sum_{j=1}^{k_c}R(\pi_i[j])\lambda_j=\frac{1}{m}\sum_{i=1}^m\sum_{j=1}^{k_c}R(\pi_i[j])p_j\\
    &=\frac{1}{m}\sum_{d\in D(q)} R(d)\big(\sum_{i=1}^{m} \sum_{j=1}^{k_c}  p_{j}\mathbbm{1}_{\pi_i[j]==d}\big)=\frac{1}{m}\sum_{d\in D(q)} R(d) E_{@k_c(d)}
    \end{split}
\vspace{0pt}
\end{equation*}
where $\pi_i[j]$ indicates the $j^{th}$ item in ranklist $\pi_{i}$, $\mathbbm{1}$ is an indicator function, $p_j$ is the $j^{th}$ rank's examining probability, $\lambda_j$ is the weight put on rank $j$. We follow \citet{singh2018fairness} to set  $\lambda_j=p_j$. $E_{@k_c(d)}=\sum_{i=1}^{m} \sum_{j=1}^{k_c}  p_{j}\mathbbm{1}_{\pi_i[j]==d}$, is item $d$'s exposure at top $k_c$ ranks. To maximize $\textit{DCG}@k_c$, we should let item of greater $R$ get more exposure, i.e., greater $E_{@k_c(d)}$. 
When items' exposure is fixed (e.g., $\alpha=1.0$ in Eq.~\ref{eq:strictFairness}), it is important to follow a greedy selection strategy to let item of greater $R$ fulfill its exposure quota at the highest ranks as we can if we think (N)$DCG$ at higher ranks are more important. In line 10 and line 11 of Algorithm~\ref{algo:fair_exp_group}, VerFair just follows the greedy selection strategy to prioritize allocating top ranks' exposure first, i.e., finish allocating  all consumers' $i^{th}$ rank before the ${(i+1)}^{th}$ rank. VerFair follows the greedy selection strategy that can maximize top ranks $DCG$, so it can reach better (N)DCG at top ranks.

\subsubsection{Proof for Minimum Exposure Guarantee}. 
\label{sec:proof_minimum}
In this section, we discuss the exposure allocation error bound between $Quota$ (the minimum exposure) and the actual allocated exposure $\Tilde{E}$ in Algorithm~\ref{algo:fair_exp_group}, i.e, $|\Tilde{E}(G)-Quota(G)|$. In the allocation phase, there exist two possible scenarios in line 12-16 of Algo.~\ref{algo:fair_exp_group}: 

\textit{Scenario 1: There exists a $(c^*,r^*)$ pair where $H \cap F(u_{c^*}) =  \varnothing$}. If this happens, $( c^*,k)$ will also have $H \cap F(u_{c^*}) =  \varnothing$ since the size of $H$ and $u_{c^*}$ monotonically decrease for lower rank of the same user. 
As $F(u_{c^*})$ is the set of unselected items for current user, we know that there are at least $n-k$ items in $F(u_c)$, i.e., $|F(u_{c^*})|\ge n-k$.  If $H\cap F(u_{c^*}) =  \varnothing$, those $n-k$ items should not be in $H$.  In other words, there are at least $|D|-k$ items whose corresponding group satisfies $Quota(G)-\Tilde{E}(G)< p_{c^*,k}$, i.e., the error is less than $p_{c^*,k}$. As we assume that $k<<|D|$ and $p_{c^*,k}<1$ , we still claim VerFair can guarantee exposure required by $Quota(G)$.

\textit{Scenario 2: $H \cap F(u_{c}) = \varnothing$ does not happen}. In other words,  line 16 in Algorithm~\ref{algo:fair_exp_group} never happens. Considering line 12 and line 20 in Algorithm~\ref{algo:fair_exp_group}, $Quota(G)-\Tilde{E}(G)\geq 0 \: \forall G \in \mathcal{G}$. And we know that  $\sum_{G \in \mathcal{G}} Quota(G)=\sum_{G \in \mathcal{G}} \Tilde{E}(G)$, i.e., the actual allocated total exposure should equal the sum of quota. The only situation is that $Quota(G)=\Tilde{E}(G) \: \forall G \in \mathcal{G}$. In other words, all groups get the exact exposure according the $Quota$.

In the above analysis, we have showed the accuracy of minimum exposure guarantee in the allocation phase. And we know that items are in lower ranks in the allocation phase and those items can only be put to higher ranks in the Appending phase and the Re-sorting phase (see examples in Tab.~\ref{tab:toy start pts rec list}). Being put at higher ranks will make minimum exposure \textit{Quota} better guaranteed. So the accuracy in the allocation phase still holds in the Appending phase and the Re-sorting phase.

\section{Experimental Setup and Results}
\label{sec:result_analysis}
In this section, we will introduce our experimental settings. Implementations will be available online\footnote{\url{https://github.com/Taosheng-ty/sigirAP-VerFair.git}.}.

\subsection{Experimental Setup}
We walk through the detailed experimental setup in this section. 

\subsubsection{Dataset and Preprocessing.} 

\begin{table}[t]
    \centering
    \caption{A detailed statistics of the datasets we use. Individual fairness datasets are used for evaluating methods for individual fairness and they don't contain any groups.}
    \vspace{-5pt}
    \begin{tabular}{l c c c} \toprule
        Datasets & \multicolumn{3}{c}{Statistics}\\ \hline
        &\#Consumers & \#Items & \#Groups\\ \hline 
        \textit{Individual fairness datasets} & \\ \hline
        Yahoo! R3  & 15,400  &1,000  &-- \\ 
        Google Local Ratings & 11,172 &855 &-- \\ \hline
        \textit{Group fairness datasets} & \\ \hline
        Movielens-Groups & 10,000  &100  & 5\\ \bottomrule
    \end{tabular}
    \label{tab:statistics}
\end{table}

The statistics of the three prepossessed datasets are shown in Table \ref{tab:statistics}.
For the individual fairness setting, we use the Google Local Ratings dataset\footnote{\href{https://cseweb.ucsd.edu/~jmcauley/datasets.html}{https://cseweb.ucsd.edu/~jmcauley/datasets.html}}\cite{he2017translation} and Yahoo! R3 datasts\footnote{\href{https://webscope.sandbox.yahoo.com/catalog.php?datatype=r}{https://webscope.sandbox.yahoo.com/catalog.php?datatype=r}}.
In order to use them in our experiment, we need to fill out the missing customer-item pair relevance in the two datasets.
\citet{patro2020fairrec} already use Matrix Factorization to fill out the Google Local Ratings dataset and can be directly downloaded here\footnote{ \href{https://github.com/gourabkumarpatro/FairRec\_www\_2020}{https://github.com/gourabkumarpatro/FairRec\_www\_2020}.}. 
For Yahoo! R3 datast, following \cite{patro2020fairrec}, we randomly sampled $1\%$ data to learn a relevance prediction model and predict all the missing customer-item pair relevance scores.
Specifically, we use SVD algorithm from Surprise Library\footnote{http://surpriselib.com/} with learning rate of 5e-3, L2 reguarization coefficient of 2e-2 and 100-d latent factors. The relevance scores are derived after 20 training iterations.
Based on the estimated consumer-item pair relevance, we construct and evaluate ranklists for consumers.

For the group fairness setting, we adopt Movielens-Groups dataset preprocessed from MovieLens Datasets (20M) by \cite{morik2020controlling}.
The missing consumer-item pair relevance are already filled out and made public\footnote{ \href{https://github.com/MarcoMorik/Dynamic-Fairness.git}{https://github.com/MarcoMorik/Dynamic-Fairness.git}}.
It contains 10,000 users and 100 movies from 5 companies/providers. 
Following the group fairness setting in \cite{morik2020controlling}, movies are grouped according to their producer companies; and there are in total 5 groups/providers. 
This partition matches our definition of provider-side fairness to fairly allocate exposure to providers.

We should note that movies are not grouped by sensitive attributes (e.g., gender, religion, or ethnicity) since fairness caused by sensitive attributes is not our main focus in this work. 
Instead of protecting candidates/candidate groups with sensitive attributes, we consider the amortized fairness principle~\cite{singh2018fairness,biega2018equity}, where candidates of similar relevance should get similar exposure. Extending our work to fairness concerning sensitive attributes is traightforward and we leave for future works.

\subsubsection{Task Definition.}
Given item set $\mathcal{D}$, consumer set $\mathcal{U}$, the presented ranklist length $k$, the consumer-item pair relevance $R(d,u |\forall d\in \mathcal{D},\forall u\in \mathcal{U})$,  our task is to generate $|\mathcal{U}|$ ranklists of length $k$. 
In other words, we will generate 15400, 11172, and 10000 ranklists for the three datasets in Table~\ref{tab:statistics} respectively. 
In our experiment, $k$ is set to 10 as default. 
The goal of the ranklist construction is to achieve better ranking relevance given the same fairness. 
Based on the consumer-item pair relevance, we use NDCG~\cite{jarvelin2002cumulated} to evaluate ranking relevance. 
To evaluate fairness, we use fairness definition in Equation (\ref{eq:fairness}). 
Note that in this work, we only focus on a post-processing setting where personal relevance is already estimated. 
As for how to get the relevance estimation, there have been many existing algorithms \cite{aciar2007informed,bao2014topicmf,ling2014ratings,mcauley2013hidden,tan2016rating,xu2022reinforcement,yang2022can,yang2023mitigating,ai2021unbiased}.


\subsubsection{Position Bias.} 
Following the experiment setting in \cite{morik2020controlling}, we use the position-biased model (PBM \cite{chuklin2015click}) to model consumer's examination behavior. 
In PBM, the probability that a consumer examines an item only depends on its position. 
We adopt the discount function of NDCG as the consumer's examination probability. 
For the $i^{th}$ rank in simulation, the examination probability is $p_i=\big(\frac{1}{log_2(1+i)} \big)^\eta$,
where $\eta$ indicates the severity of position bias. 
The greater $\eta$ is, the more exposure consumers put on top ranks. 
In our experiment, we adopt the same setup as \cite{biega2018equity,morik2020controlling} where we assume examination probabilities are already known and all consumers follow the same position bias. 
As for how to estimate the examination probabilities, many mature methods \cite{ai2018unbiased,wang2018position,agarwal2019estimating,radlinski2006minimally} have been proposed, which is beyond the scope of this paper.

\subsubsection{Baseline Methods.}
We summarize the methods we will compare in this paper as follows:

\begin{itemize}[leftmargin=*]
    \vspace{-5pt}
    \item \textbf{Top-k}: Select top-$k$ items according to personal relevance.
    \item \textbf{Random-k}: Randomly select k items.
    \item \textbf{PR-k}, (Poor-k): A fair algorithm which dynamically selects $k$ most under-exposed items ($\arg_{topK} Quota(d|\alpha=1)-E(d) $) for the current consumer. 
    \item \textbf{ILP-Aver}: Integer linear programming method of individual fairness proposed by \citet{biega2018equity}, which considers the average relevance. Compared with other methods, it only ranks items in a non-personalized way. Tradeoff parameter range lies in [0.0,1.0].
    \item \textbf{ILP-Pers}: Integer linear programming method of individual fairness proposed by \citet{biega2018equity} which considers personal relevance. Tradeoff parameter range lies in [0.0,1.0].
    \item \textbf{FairCo}: Amortized fairness method at both individual level and group level proposed by \citet{morik2020controlling}. Tradeoff parameter range lies in $[0.0,+\infty]$. We adopt $[0.0,1000]$ in our experiments.
    \item \textbf{FairRec}: Unfair method proposed by \citet{patro2020fairrec} to only ensure equal frequency for items, thus not a amortized fairness method. Tradeoff parameter range lies in [0.0,1.0].
    \item \textbf{VerFair(Ind)}: Our method of individual fairness mode. Tradeoff parameter range lies in [0.0,1.0].
    \item \textbf{VerFair(Group)}: Our method of group fairness mode. Tradeoff parameter range lies in [0.0,1.0].
\\
\vspace{-15pt}
\end{itemize}
Among the above methods, only FairRec and our method VerFair construct ranklists in vertical way while all other methods follow a horizontal setup. 
All methods above except Top-k, Random-k, and PR-k have tradeoff parameters to tradeoff between ranking relevance and fairness. 
We have adjustable tradeoff parameter to make balance the weight between relevance and fairness. 
For example, when the tradeoff parameter is set $0$, the minimum value, VerFair(Ind), FairCo, ILP-Pers,VerFair(Group), and FairRec will degenerate to Top-k methods, where they only care about relevance and ignore fairness. 
For methods Top-k, Random-k, and PR-k, they don't have tradeoff parameters and can't adjust the weight between fairness and ranking relevance.

\subsection{Experimental Results}

\begin{figure*}[h]
    \centering
    \begin{subfigure}[b]{0.21\textwidth}
  \includegraphics[width=\textwidth]{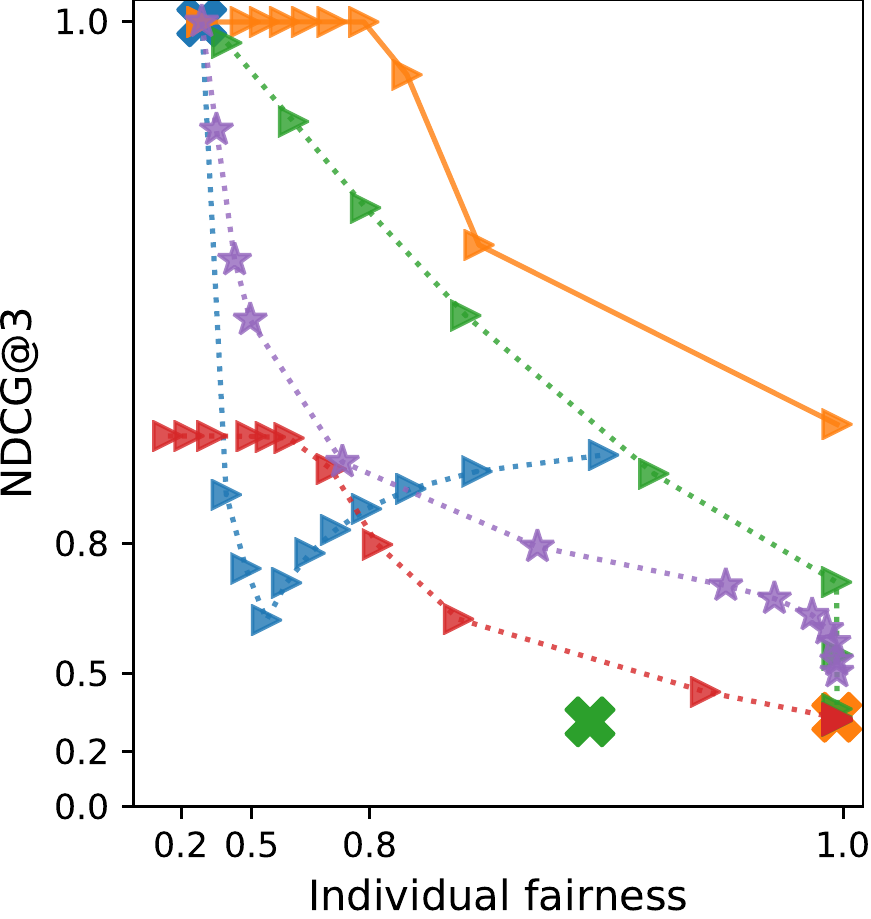}
    \caption{ Yahoo! R3,  $\eta=0$}
      \vspace{0pt}
    \label{fig:tradeoff_11}
    \end{subfigure}
    \hfill
    \begin{subfigure}[b]{0.21\textwidth}
  \includegraphics[width=\textwidth]{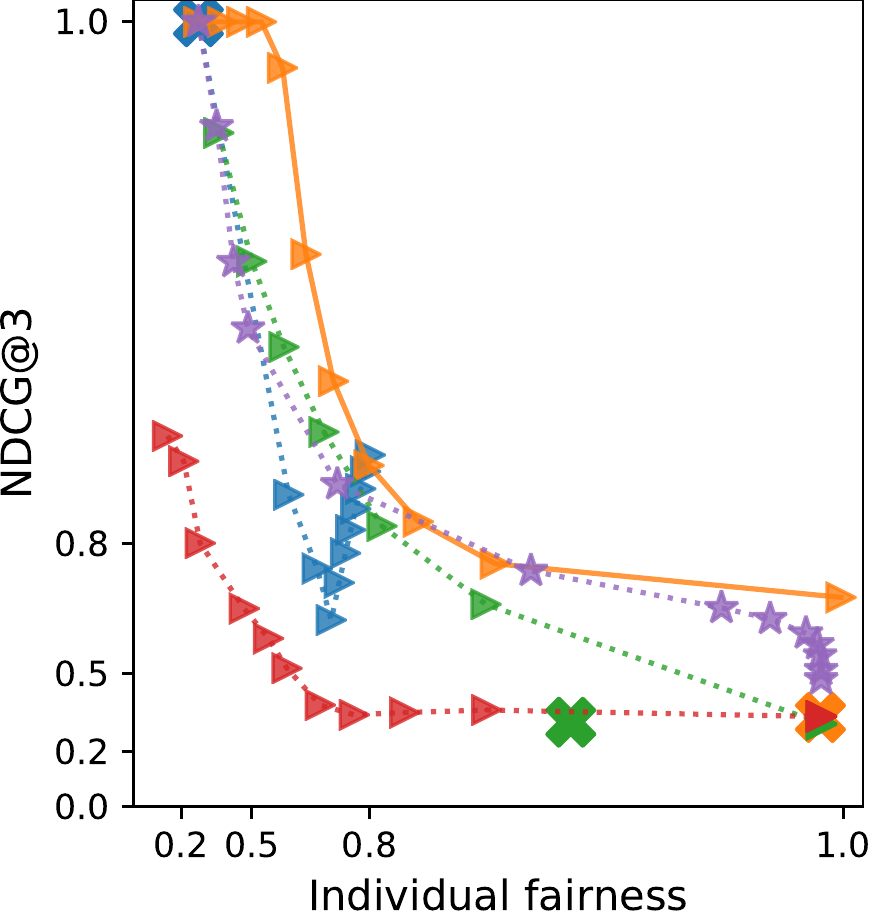}
    \caption{ Yahoo! R3,  $\eta=2$}
    \vspace{0pt}
    \label{fig:tradeoff_12}
    \end{subfigure}
    \hfill
    \begin{subfigure}[b]{0.21\textwidth}
  \includegraphics[width=\textwidth]{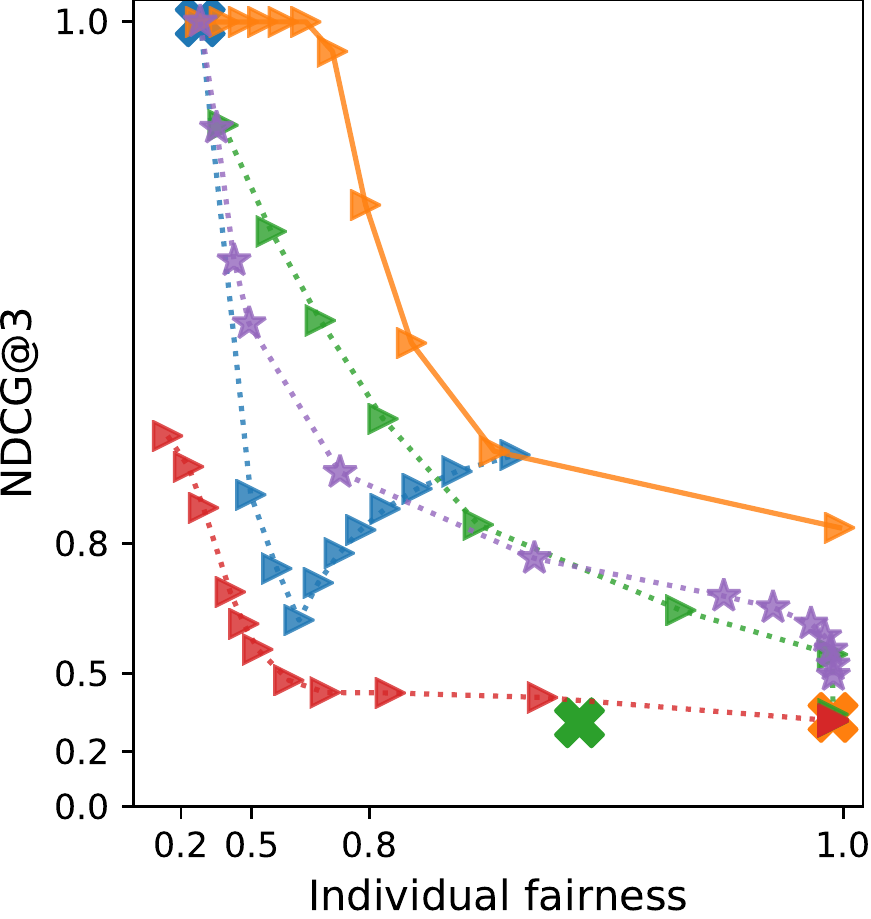}
    \caption{Yahoo! R3,  $\eta=1$}
    \vspace{0pt}
    \label{fig:tradeoff_13}
    \end{subfigure}
    \hfill
    \begin{subfigure}[b]{0.21\textwidth}
  \includegraphics[width=\textwidth]{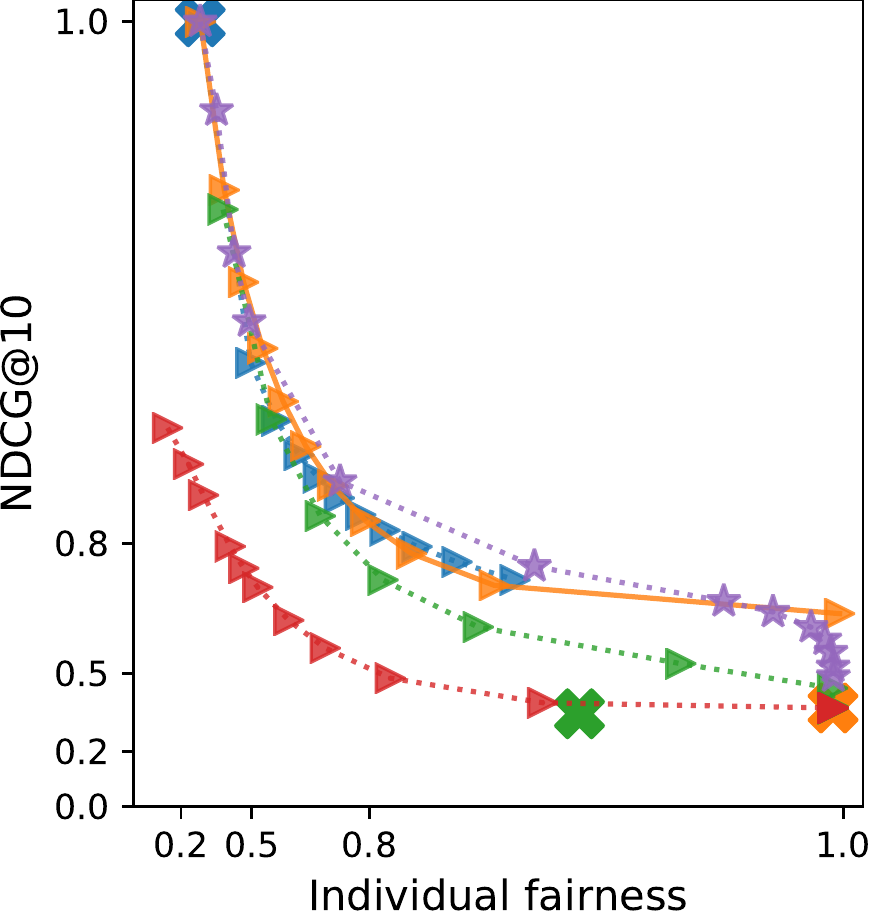}
    \caption{ Yahoo! R3, $\eta=1$}
    \vspace{0pt}
    \label{fig:tradeoff_14}
    \end{subfigure}
    \hfill
    \begin{subfigure}[b]{0.11\textwidth}
    \includegraphics[width=\textwidth]{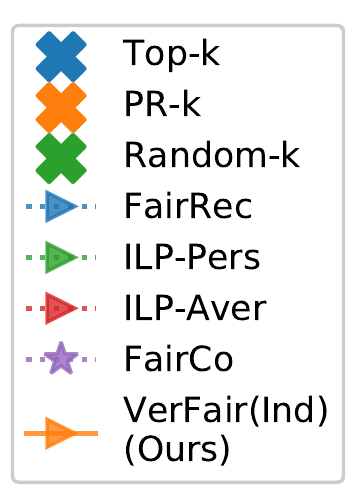}
\vspace{0pt}
\label{fig:tradeoff_15}
\end{subfigure}

      \begin{subfigure}[b]{0.21\textwidth}
  \includegraphics[width=\textwidth]{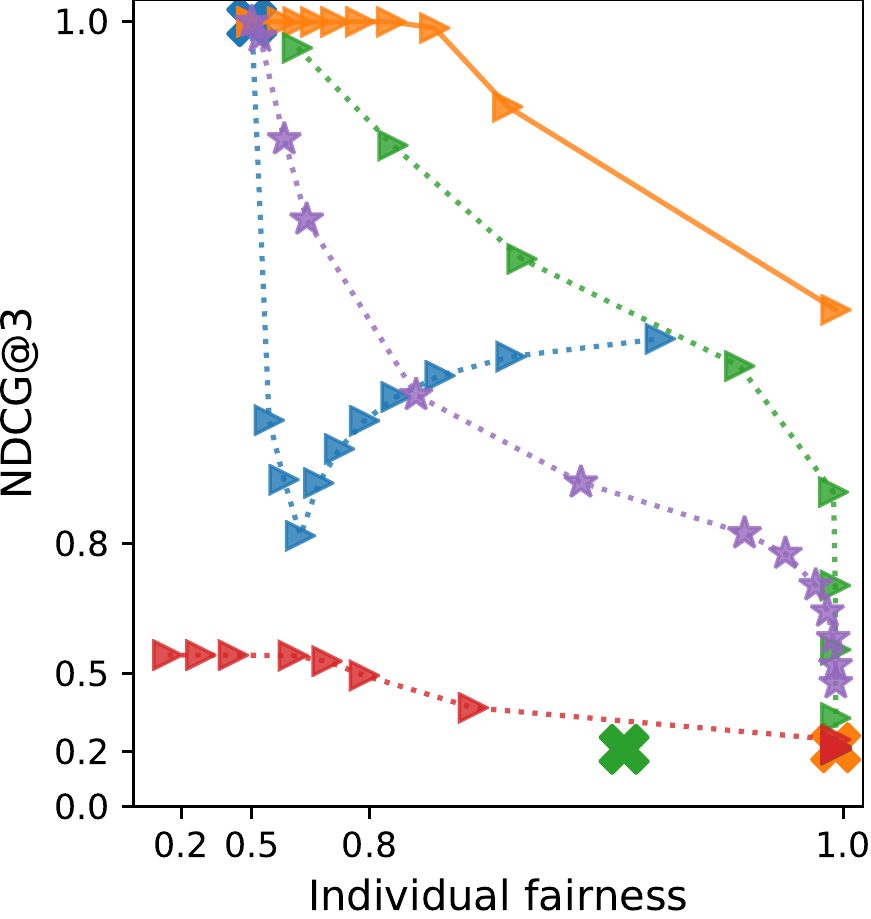}  
        \caption{ Google Local ,  $\eta=0$}
          \vspace{0pt}
        \label{fig:tradeoff_21}
    \end{subfigure}
  \hfill
        \begin{subfigure}[b]{0.21\textwidth}
  \includegraphics[width=\textwidth]{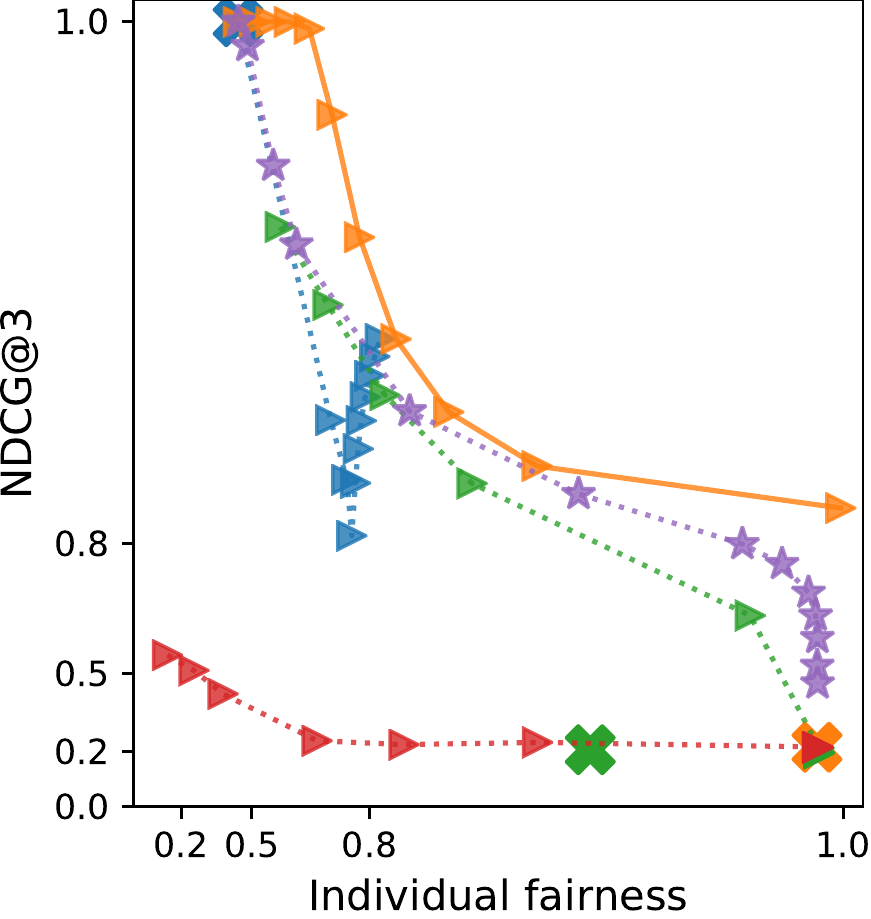}
          \caption{Google Local ,  $\eta=2$}
          \vspace{0pt}
        \label{fig:tradeoff_22}
    \end{subfigure}
    \hfill
          \begin{subfigure}[b]{0.21\textwidth}
  \includegraphics[width=\textwidth]{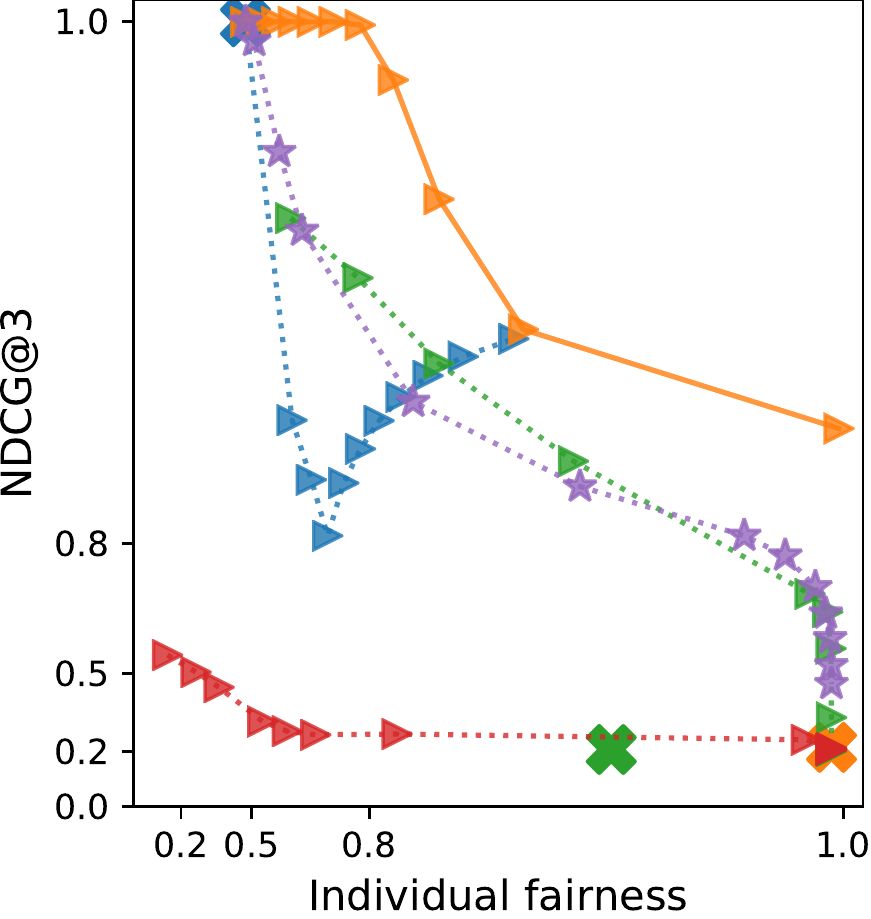}
          \caption{ Google Local ,  $\eta=1$}
          \vspace{0pt}
        \label{fig:tradeoff_23}
    \end{subfigure}
    \hfill
          \begin{subfigure}[b]{0.21\textwidth}
  \includegraphics[width=\textwidth]{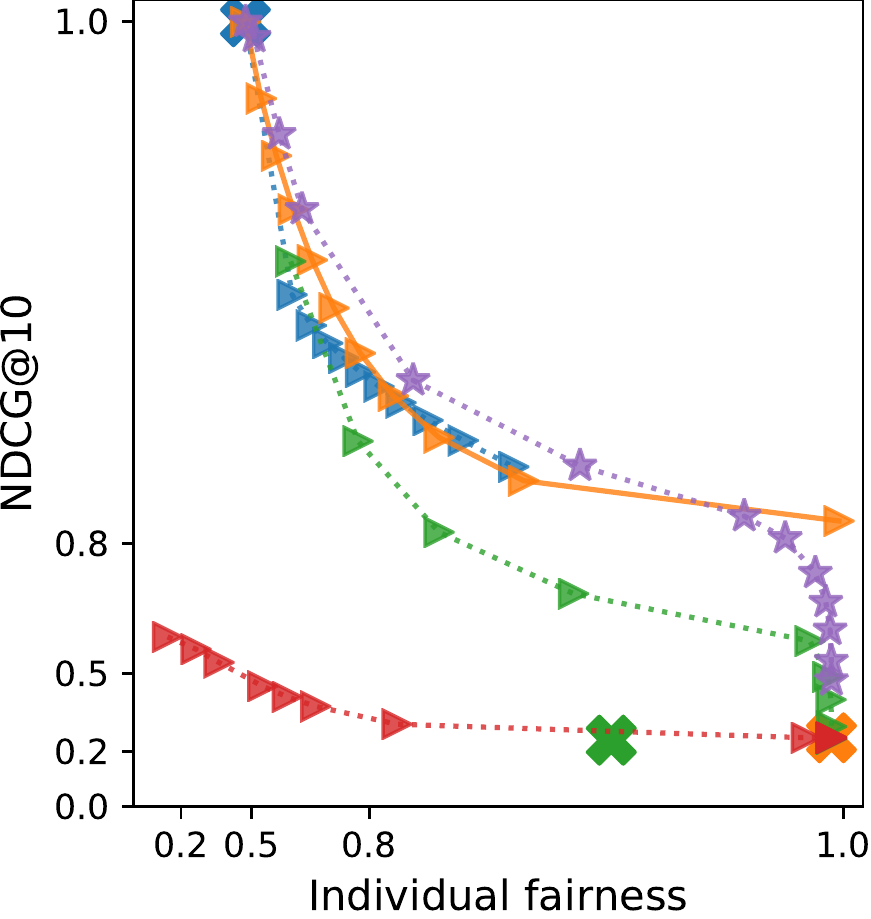}
          \caption{ Google Local, $\eta=1$}
          \vspace{0pt}
        \label{fig:tradeoff_24}
    \end{subfigure}  
      \hfill
          \begin{subfigure}[b]{0.11\textwidth}
  \includegraphics[width=\textwidth]{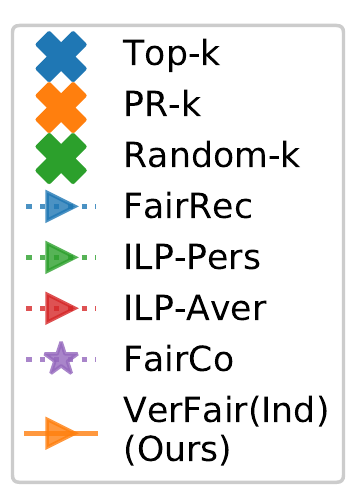}
        \vspace{0pt}
        \label{fig:tradeoff_25}
    \end{subfigure}

    \begin{subfigure}[b]{0.21\textwidth}
  \includegraphics[width=\textwidth]{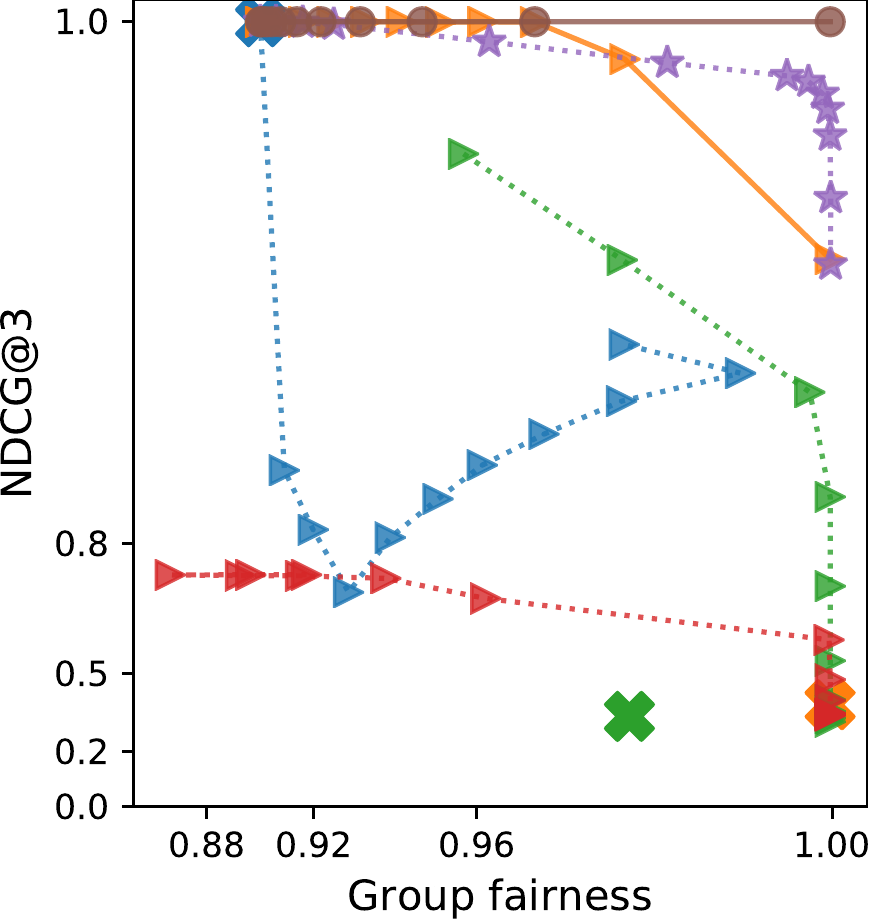}  
    \caption{Movielens-Groups,  $\eta=0$}
    \vspace{0pt}
    \label{fig:tradeoff_31}
    \end{subfigure}
    \hfill
    \begin{subfigure}[b]{0.21\textwidth}
  \includegraphics[width=\textwidth]{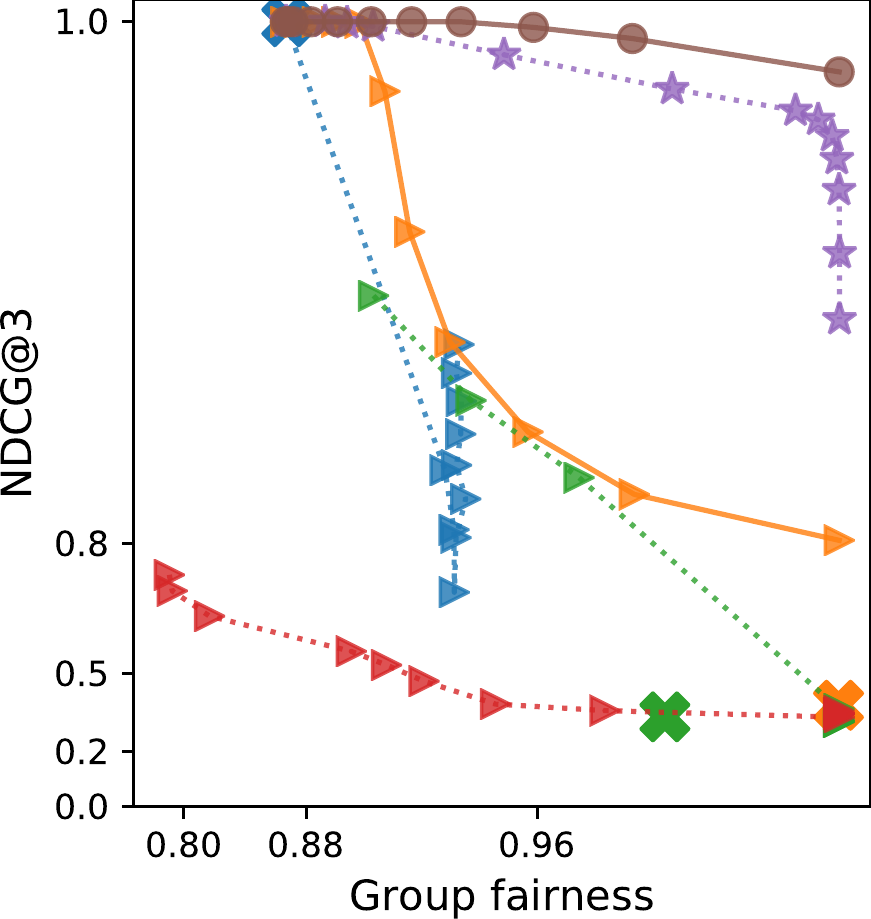}
    \caption{ Movielens-Groups,  $\eta=2$}
    \vspace{0pt}
    \label{fig:tradeoff_32}
    \end{subfigure}
    \hfill
    \begin{subfigure}[b]{0.21\textwidth}
  \includegraphics[width=\textwidth]{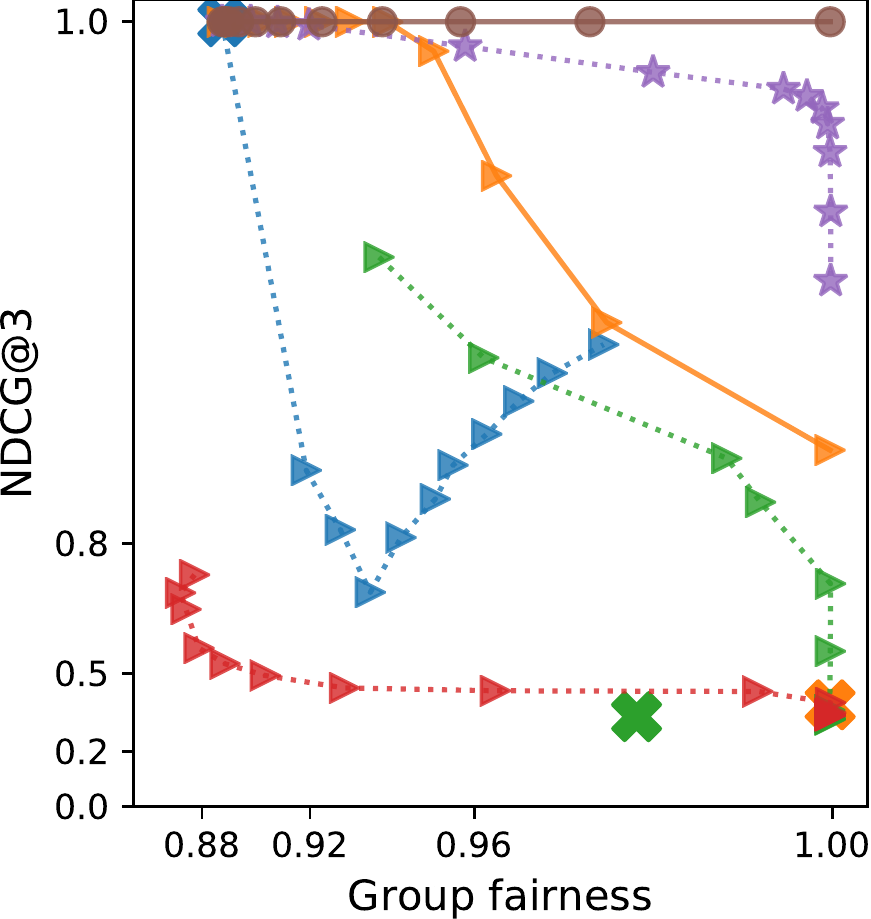}
    \caption{ Movielens-Groups,  $\eta=1$}
    \vspace{0pt}
    \label{fig:tradeoff_33}
    \end{subfigure}
    \hfill
    \begin{subfigure}[b]{0.21\textwidth}
  \includegraphics[width=\textwidth]{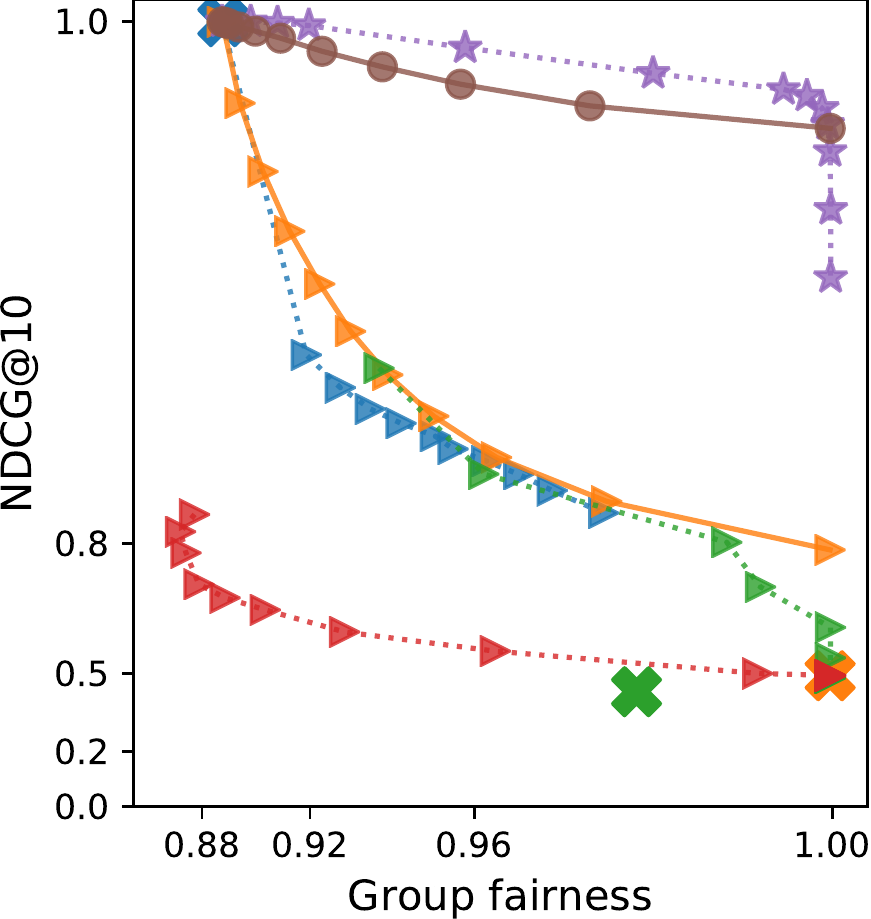}
    \caption{ Movielens-Groups, $\eta=1$}
    \vspace{0pt}
    \label{fig:tradeoff_34}
    \end{subfigure}  
    \hfill
    \begin{subfigure}[b]{0.11\textwidth}
    \includegraphics[width=\textwidth]{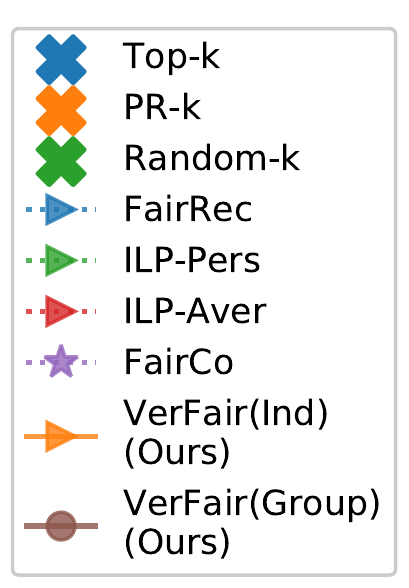}
    \vspace{0pt}
    \label{fig:tradeoff_35}
    \end{subfigure}      

  \vspace{-5pt}
  \caption{Tradeoff between fairness (x-axis) and NDCG (y-axis) for ranklists of length $k=10$. Our methods VerFair(Ind) and VerFair(Group)  are shown in solid lines, while others are dotted lines. Result curves start from top left to bottom right when $\alpha$ increases, i.e., caring more about fairness and sacrificing NDCG. Curves lies in top right show better tradeoff. Top-k, PR-k and Random-k don't have tradeoff parameters so they are points. Note that x-axis is not $\alpha$. } 
  \label{fig:tradeoff}
  \vspace{-10pt}
\end{figure*}

\subsubsection{How does VerFair perform compared to baselines?}
\label{sec:tradeoff}
Figure \ref{fig:tradeoff} shows the tradeoff curves between NDCG and fairness for different methods after we iterate the tradeoff parameters. As there are no tradeoff parameters for Top-k, PR-k and Random-k, their performances are actually points in the graphs. Among the methods, Top-k is the best method for NDCG, while PR-k is the best method for fairness. All amortized fairness methods, i.e., VerFair, ILP-Aver, ILP-Pers, FairCo, and PR-k, can have fairness metrics near 1.0 when the tradeoff parameter reaches its maximum (i.e., bottom right area), which prove their effectiveness to reach fairness. For Random-k, it randomly selects k items and thus can't reach fairness. 

As is shown in Figure (\ref{fig:tradeoff_11},\ref{fig:tradeoff_12},\ref{fig:tradeoff_13}) and Figure  (\ref{fig:tradeoff_31},\ref{fig:tradeoff_32},\ref{fig:tradeoff_33}), 
our method VerFair(Ind) and VerFair(Group) significantly outperform ILP-Pers, ILP-Aver and FairCo in terms of balance between NDCG@3 and fairness under various degrees of position bias severity, i.e., $\eta=0,1,2$. 
Given the same degree of fairness, VerFair can reach higher NDCG@3 than ILP-Pers, ILP-Aver and FairCo. 
Given the same NDCG@3,  VerFair can get fairer ranklists. ILP-Pers performs better than ILP-Aver because ILP-Pers can perform personalized ranking. 
In addition, we didn't observe a clear tradeoff between relevance and fairness in FairRec when $\eta$ is greater than 0. 
It meets our expectation since FairRec is not an amortized fairness method.

Figure \ref{fig:tradeoff_14}, \ref{fig:tradeoff_24}  and \ref{fig:tradeoff_34} show the tradeoff between NDCG@10 and fairness when ranklists evaluated at $k=10$. As we can see from the figures, our method VerFair(Ind) and VerFair(Group) show similar or slightly inferior results on long prefixes (@10) compared to FairCo. 
Given same degree of fairness, our algorithm VerFair focuses more on top ranks and puts more relevant items on top ranks. 
We believe slight compromise is unavoidable. 
Since VerFair tends to put more relevant items on top ranks, to keep the same degree of fairness, some relatively irrelevant items will be put at lower ranks. 
Thus, the advantages of our methods are more significant on top ranks than low ranks. 
An additional note is that different from NDCG, fairness doesn't need to do cutoff evaluation \cite{biega2018equity,singh2018fairness} (Eq. \ref{eq:fairness}).
Since fairness evaluation cares about exposure and it is not reasonable to ignore lower ranks' exposure even if they are small.

Since fair methods at an individual level can automatically reach group fairness, individual fairness methods VerFair(Ind), ILP-Pers, ILP-Aver also reach fairness in group fairness settings, as shown in Figure \ref{fig:tradeoff_31}, \ref{fig:tradeoff_32}, \ref{fig:tradeoff_33} and \ref{fig:tradeoff_34}. However, as individual fairness brings more constraint than group fairness, all individual fairness methods show a dramatic drop in NDCG in Movielens-Groups dataset. 

\subsubsection{Can VerFair reach fairness while maintaining good ranking quality?}
\label{sec:dist}

\begin{figure}[t]
\centering
    \begin{subfigure}[]{0.23\textwidth}
        \includegraphics[width=\textwidth,height=3.5cm]{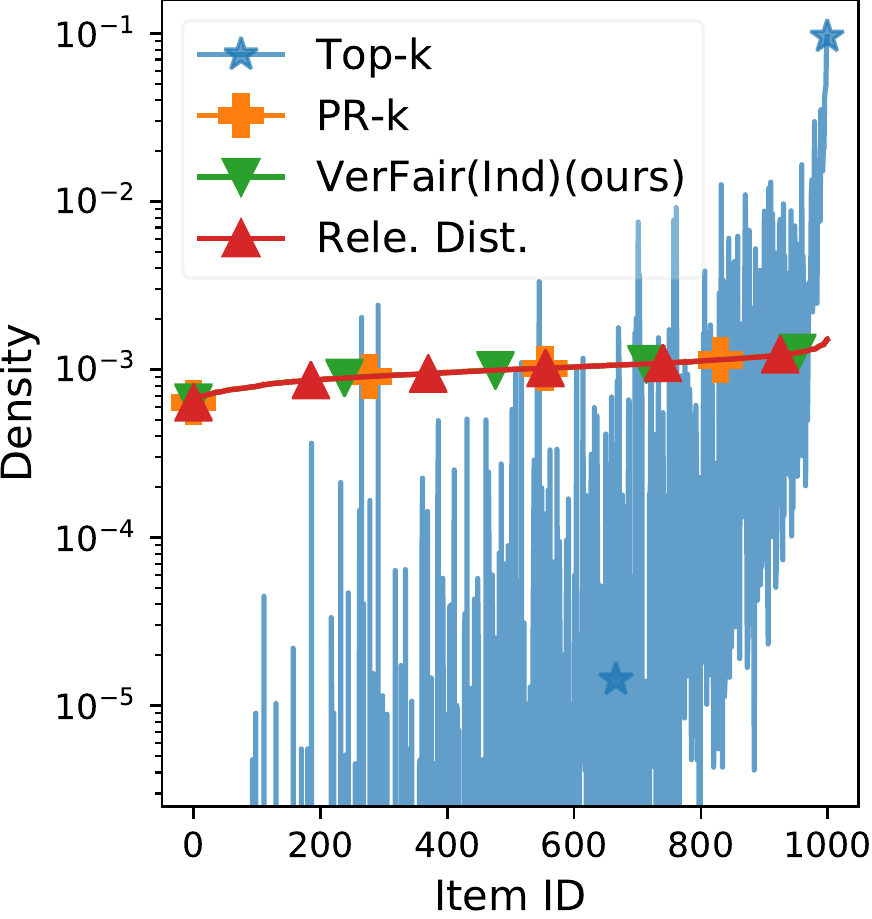}
        \caption{Yahoo! R3}
        \label{fig:dist_a}
    \end{subfigure}
    \hfill
    \begin{subfigure}[]{0.23\textwidth}
        \includegraphics[width=\textwidth,height=3.5cm]{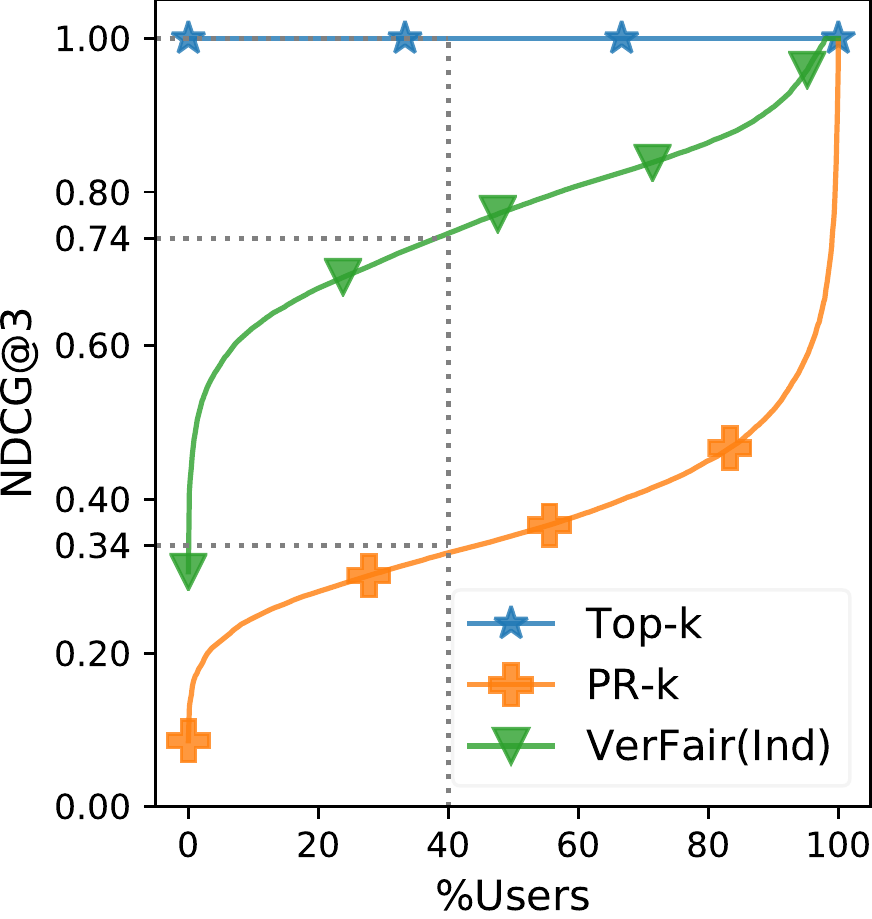}
        \caption{Yahoo! R3}
        \label{fig:dist_b}
    \end{subfigure}
\vspace{-5pt}
\caption{Experimental results on Yahoo R3! dataset. (a) and (b) show the density distribution of relevance and exposure (exposure defined in Eq.\ref{eq:eq(d)} and  in Eq.\ref{eq:RE(G)}) where position bias severity $\eta=1$, ranklists length $k=10$. Exposure distribution from fair methods PR-k and VerFair ($\alpha=1.0$) match relevance distribution (i.e., PR-k and VerFair curves overlap Relevance Dist. curves), while results from unfair method Top-k don't.}
\label{fig:dist}
\end{figure}

Due to limited space, we only provide analysis on Yahoo R3! dataset. 
Figure \ref{fig:dist_a} show the density distribution of relevance and exposure by using different methods. 
The red line with triangles stands for relevance distribution of items.
The goal of amortized fairness is that exposure distribution should match the relevance distribution. 
In other words, a perfect amortized fairness ranking algorithm should produce an exposure distribution that can exactly match the line of Relevance Dist (the red lines with triangles) in Figure \ref{fig:dist_a}. 
Distributions of exposure from method PR-k and VerFair are highly overlapping with Relevance Dist.
In contrast, distributions of exposure from unfair method Top-k are dramatically different from Relevance Dist., thus showing a huge sacrifice of amortized fairness. We now take a look at  customers' satisfaction, i.e., the NDCG distribution in Figure \ref{fig:dist_b}. 
The unfair method Top-k recommends top k items according to personal relevance, so it can always reach the skyline NDCG, i.e., 1. The highest NDCG is 1 because we focus on post-processing and assume (personal) relevance is already known in this paper. 
PR-k shows a huge drop in NDCG since it only centers on  fairness of item rankings. 
In contrast, our method VerFair can achieve significantly better NDCG than PR-k while achieving similar amortized fairness. 


\subsubsection{Can VerFair guarantee minimum exposure?}
\begin{figure}[t]
\centering
\begin{subfigure}[]{0.22\textwidth}
    \centering
    \includegraphics[width=\textwidth,height=3.5cm]{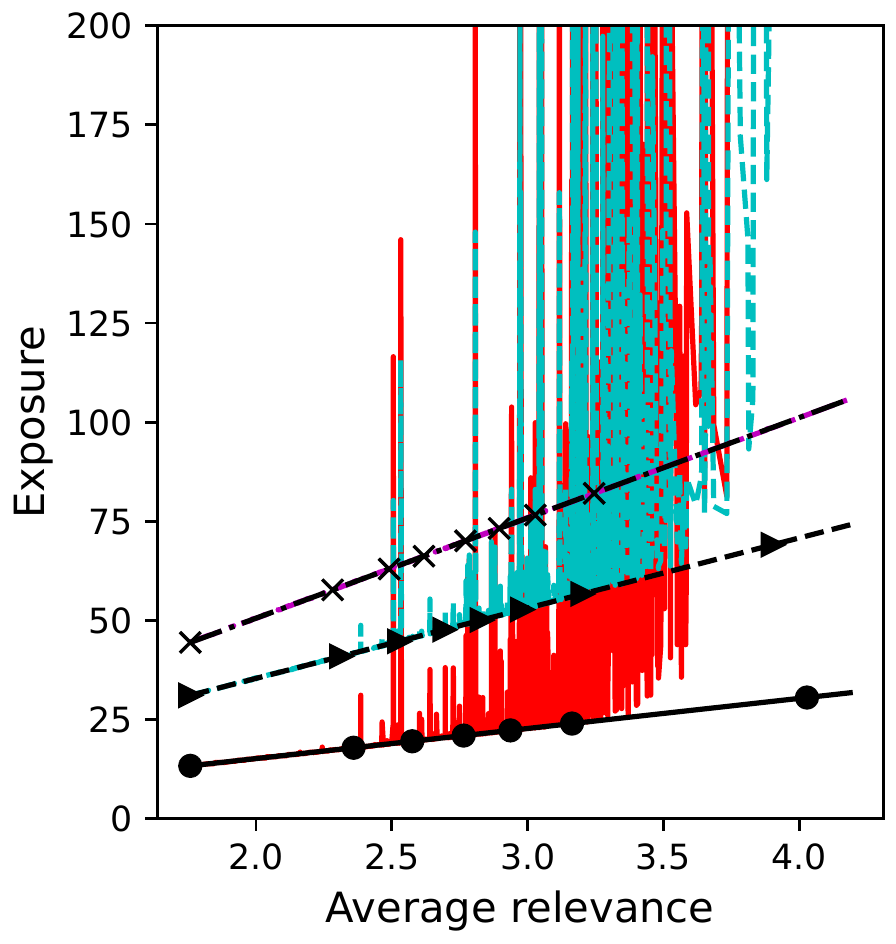}
    \caption{Exposure-Relevance}
    \label{fig:part_expo}
\end{subfigure} \hfill
\begin{subfigure}[]{0.22\textwidth}
    \centering
    \includegraphics[width=\textwidth,height=3.5cm]{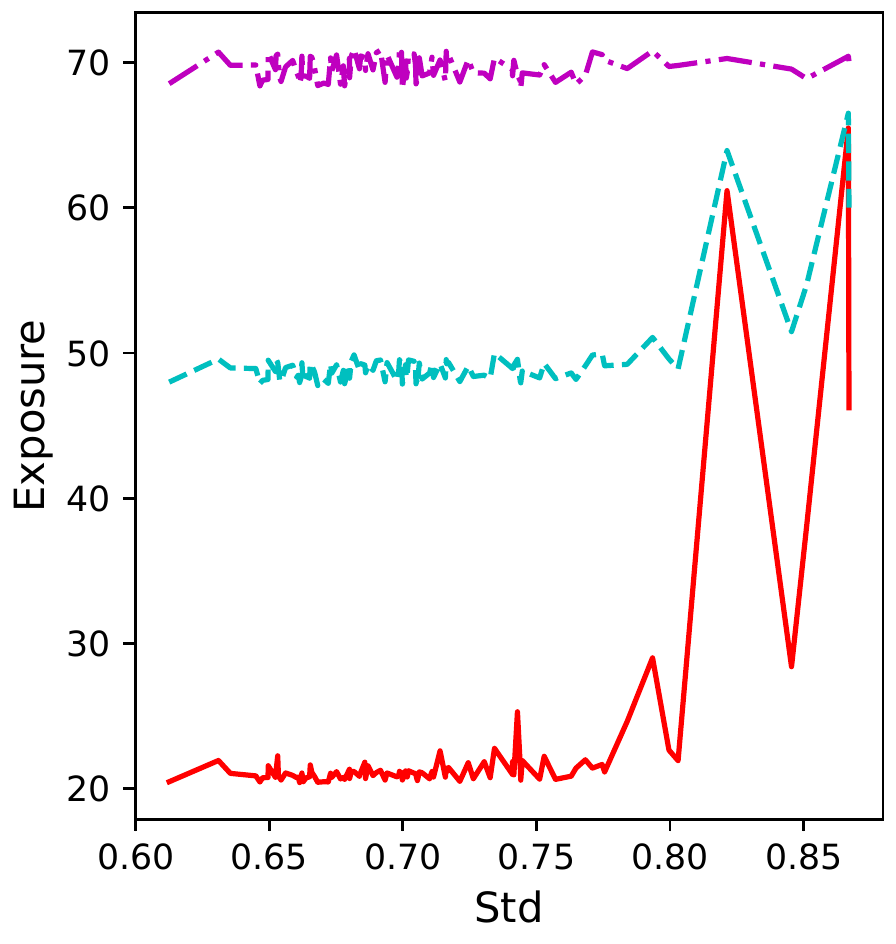}
    \caption{Exposure-Std}
    \label{fig:part_std}
\end{subfigure}
\begin{subfigure}[]{0.44\textwidth}
    \centering
    \includegraphics[width=\textwidth]{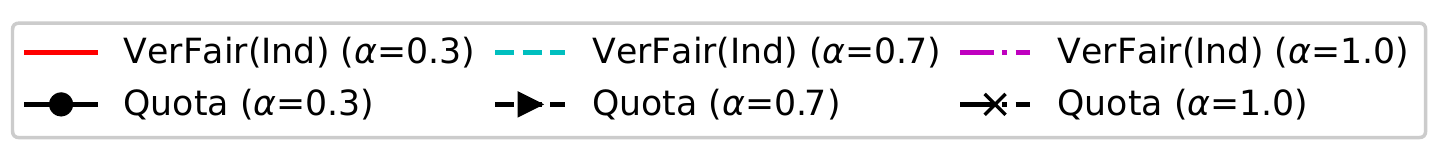}
\end{subfigure}
\vspace{-5pt}
\caption{
(a) Exposure distribution along average relevance when $\alpha$ are 0.3, 0.7 and 1.0 respectively. 
Quota($\alpha$) indicates the relevance induced minimum exposures (see Eq. (\ref{eq:quota_ind})). 
(b) Exposure distribution along the deviation for items of similar average relevance (within range $[2.7,2.8]$). 
Items of greater deviation are more tailored to customers. 
Flat curves indicate discouraging personalization. $k=10,\eta =1.0$}
\end{figure}

\label{sec:exposure_guarantee}
Due to the limited space, we use Yahoo R3! experiments results as an example. As shown in Figure \ref{fig:part_expo},   
VerFair can guarantee the minimum exposure quota under different $\alpha$ (select $\alpha$ as 0.3, 0.7, 1.0 as examples) 
since result exposure distribution curves of VerFair($(\alpha)$) are above the $\textit{Quota}(\alpha)$. 
The two curves overlap when $\alpha=1.0$ because all exposure are used to calculate the minimum exposure quota (Eq.\ref{eq:quota_ind}). 
Also, with VerFair method, items still have a chance to gain more exposure than their quota if they have better personalization to their target users. 
To demonstrate it, we select items of average relevance within $[2.7,2.8]$ for Yahoo R3!. 
We choose this interval as it has the largest number of items. 
Since the interval is narrow, we can assume those selected items have the same average relevance. 
Items with greater standard deviation (Std) typically do a higher personalization.  
As shown in Figure \ref{fig:part_std}, items usually don't get extra exposure when amortized fairness is strictly maintained, i.e. $\alpha=1.0$. 
When $\alpha \leq 1$, we can see a clear positive correlation between std and exposure. 
Such correlation means our method actually promotes items that are personalized for specific users but not all users.

\subsubsection{How does VerFair perform compared to baselines in terms of computational efficiency?}
\label{sec:efficiency}
In order to show the efficiency of VerFair, we test the average time (seconds) to generate 1k ranklists, which is shown in Table \ref{tab:speed}. As shown in the table, ILP methods are NP-complete [9], both ILP-Aver and ILP-Pers are time-consuming and not likely to satisfy the requirement of large-scale ranking services in practice. While for other methods, theoretically, FairRec, VerFair and FairCo have time complexity as $O(m\times n\times k)$ when there are $m$ users, $n$ items, and the length of ranklist is $k$. Among them, FairRec is not an amortized fairness method. FairCo is originally designed for group fairness and is efficient in group settings. However, its efficiency drops when we apply it on individual levels. Empirically, VerFair has better computational efficiency than all the baselines. More comparisons of those methods can be found in related works. All the experiments are conducted on Intel(R) Xeon(R) CPU E5-2640 (2.4GHz) and 252G of memory.
\begin{table}[t]
\vspace{-10pt}
\centering
    \caption{Average time (seconds) to generate 1k ranklists for different methods. k=10,$\eta=1$}
    \vspace{-5pt}
    \centering
    \resizebox{0.85\columnwidth}{!}
    {
    \begin{tabular}{c c c c}\toprule
        \multirow{2}{*}{Methods} & \multicolumn{3}{c}{Datasets}\\ \cline{2-4}
        &{Yahoo R3!} &{Google Local}&{Movielens\_Groups}\\ 
        \hline
        VerFair(Ind)&   0.2	&0.2&0.1 \\ 
        VerFair(Group)&{--}  &{--}&0.1 \\ \midrule
        ILP-Aver&  44.2&46.4&	46.4 \\ 
        ILP-Pers& 147.2&98.7  &	57.5 \\
        FairCo&  4.7&3.7	&	0.1 \\     
        FairRec&  1.6&1.5	&	0.3\\ \bottomrule  
    \end{tabular}
    }
    \label{tab:speed}
\end{table}

\section{Conclusion and Future Work}
\label{sec:summary}
We propose VerFair with the aim of reaching a better balance between fairness and ranking relevance.
With a novel vertical allocation strategy, VerFair can effectively amortize exposure and achieve amortized fairness at both the individual level and the group level. 
In the future, we will extend current work to explore further the
dynamic interactions among consumers, items, and platforms.
\begin{acks}
This work was supported  by the School of Computing, University of Utah. Any opinions, findings, conclusions, or recommendations expressed in this material are those of the authors and do not necessarily reflect those of the sponsor.
\end{acks}
\newpage
\bibliographystyle{ACM-Reference-Format}
\bibliography{mybib}
\balance
\pagebreak

\setcounter{equation}{0}
\setcounter{figure}{0}
\setcounter{table}{0}
\setcounter{page}{1}
\setcounter{section}{0}
\renewcommand{\thesection}{S-\Roman{section}}
\makeatletter
\renewcommand{\theequation}{S\arabic{equation}}
\renewcommand{\thefigure}{S\arabic{figure}}
\renewcommand{\thetable}{S\arabic{table}}
\renewcommand{\bibnumfmt}[1]{[S#1]}
\renewcommand{\citenumfont}[1]{S#1}

\end{document}